\documentclass[12pt]{article}
\pdfoutput=1
\usepackage[a4paper,bottom=4.2cm,top=1cm,head=3cm,width=18cm,dvipdfm]{geometry}
\evensidemargin=\oddsidemargin

\usepackage[dotinlabels]{titletoc}
\usepackage{titlesec}
\usepackage{ulem}

\usepackage{xcolor}
\usepackage{mathrsfs}
\usepackage{amsmath}
\usepackage{amssymb}
\usepackage{bm}
\usepackage{amsfonts}
\usepackage{subfigure}
\usepackage{feynmp}
\usepackage{extarrows}
\usepackage{slashed}
\usepackage{graphicx}
\usepackage{dcolumn}
\usepackage{verbatim}
\usepackage{here}
\usepackage{multirow} 
\allowdisplaybreaks[4]
\usepackage{indentfirst}
\usepackage{isodateo}

\definecolor{gesfpurple}{rgb}{0.47,0.19,0.42}

\definecolor{gesflanse}{rgb}{0.00,0.50,0.50}

\definecolor{gesfblue}{rgb}{0.08,0.42,0.76}

\definecolor{gesfred}{rgb}{1,0,0}

\definecolor{gesfwhite}{rgb}{1,1,1}

\definecolor{gesfblack}{rgb}{0,0,0}

\numberwithin{equation}{section}
\renewcommand{\thefootnote}{\arabic{footnote}}
\usepackage{isodateo}
\usepackage[CJKbookmarks=true, bookmarksnumbered=true,bookmarksopen=true,]{hyperref}
\hypersetup{colorlinks,%
              linkcolor=blue,
              citecolor=blue,
              urlcolor=blue}

\graphicspath{{figs/}}

\newcommand{\be}{\begin{equation}}
\newcommand{\ee}{\end{equation}}
\newcommand{\bea}{\begin{eqnarray}}
\newcommand{\eea}{\end{eqnarray}}

\def\ede{\end{equation}}
\def\bga{\begin{aligned}}
\def\eda{\end{aligned}}
\newcommand{\beq}{\begin{equation}}
\newcommand{\eeq}{\end{equation}}
\newcommand{\bq}{\begin{equation}}
\newcommand{\eq}{\end{equation}}
\newcommand{\ba}{\begin{array}}
\newcommand{\ea}{\end{array}}
\newcommand{\beqa}{\begin{eqnarray}}
\newcommand{\eeqa}{\end{eqnarray}}
\newcommand{\beqs}{\begin{subequations}}
\newcommand{\eeqs}{\end{subequations}}

\def\nn{\nonumber}

\def\({\left(}
\def\){\right)}

\def\leqq{\leqslant}
\def\geqq{\geqslant}
\def\End{\end{document}}

\def\d{\text{d}}

\def\be{\beta}

\def\la{\lambda}

\def\MX{M_{\chi}^{}}

\def\End{\end{document}}

\begin{document}

 \thispagestyle{empty}
 \renewcommand{\thefootnote}{\fnsymbol{footnote}}
 \setcounter{footnote}{0}
 \titlelabel{\thetitle.\quad \hspace{-0.8em}}
\titlecontents{section}
              [1.5em]
              {\vspace{4mm} \large \bf}
              {\contentslabel{1em}}
              {\hspace*{-1em}}
              {\titlerule*[.5pc]{.}\contentspage}
\titlecontents{subsection}
              [3.5em]
              {\vspace{2mm}}
              {\contentslabel{1.8em}}
              {\hspace*{.3em}}
              {\titlerule*[.5pc]{.}\contentspage}
\titlecontents{subsubsection}
              [5.5em]
              {\vspace{2mm}}
              {\contentslabel{2.5em}}
              {\hspace*{.3em}}
              {\titlerule*[.5pc]{.}\contentspage}
\titlecontents{appendix}
              [1.5em]
              {\vspace{4mm} \large \bf}
              {\contentslabel{1em}}
              {\hspace*{-1em}}
              {\titlerule*[.5pc]{.}\contentspage}

\def\thisday{February 23, 2017}


\vspace*{-20mm}

\begin{center}
{\Large\bf Probing Flavor Structure of Cosmic Ray $\boldsymbol{e^\mp}$ Spectrum
\\[2mm]
and Implications for Dark Matter Indirect Searches}

\vspace*{8mm}

{\sc Shao-Feng Ge},$^{a,}$\footnote{Email: gesf@sjtu.edu.cn}~
{\sc Hong-Jian He},$^{a,b,c,}$\footnote{Email: hjhe@sjtu.edu.cn, hjhe@tsinghua.edu.cn}~
{\sc Yu-Chen Wang},$^{b,d,}$\footnote{Email: wang-yc15@mails.tsinghua.edu.cn}~
{\sc Qiang Yuan}\,$^{e,f,c,}$\footnote{Email: yuanq@pmo.ac.cn}

\vspace*{4mm}

$^a$\,Tsung-Dao Lee Institute and School of Physics and Astronomy,\\
Shanghai Jiao Tong University, Shanghai 200240, China
\\[1.5mm]
$^b$\,Institute of Modern Physics and Department of Physics, \\
Tsinghua University, Beijing 100084, China
\\[1.5mm]
$^c$\,Center for High Energy Physics, Peking University, Beijing 100871, China
\\[1.5mm]
$^d$\,Department of Physics, Harvard University, Cambridge, MA 02138, USA
\\[1.5mm]
$^e$\,Key Laboratory of Dark Matter and Space Astronomy,
Purple Mountain Observatory,\\
Chinese Academy of Sciences, Nanjing 210033, China
\\[1.5mm]
$^f$\,School of Astronomy and Space Science, \\
University of Science and Technology of China, Hefei 230026, China

\end{center}

\vspace*{3mm}

\begin{abstract}
\baselineskip 17pt
\noindent
Measuring high energy cosmic ray electrons/positrons (CRE) provides important
means for the dark matter (DM) indirect detection and for probing the nearby
galactic sources. In this work, we perform a systematic analysis of the
flavor structure of DM annihilations into charged leptons based on the
cosmic ray CRE spectra measured by DAMPE, Fermi-LAT, AMS-02, and CALET experiments.
We study the annihilations of possible TeV scale DM particles in a nearby subhalo,
which is proposed to explain the possible peak-like structure of the DAMPE CRE data.
We pay special attention to the possible non-resonant excess (besides the possible
peak-like structure) and demonstrate that such non-resonant excess can mainly arise
from the decay of muons produced by the DM annihilations in the subhalo.
With these we study the flavor composition of the lepton final states from
DM annihilations $\,\chi\chi\!\to  e^+e^-\!,\,\mu^+\mu^-\!,\,\tau^+\tau^-$\,
by fitting the CRE data.
We demonstrate that decays of the final states $\mu^+ \mu^-$ and $\tau^+ \tau^-$
can provide the non-resonant excess,
while the peak excess arises from the $e^+ e^-$ final state.
We further analyze the constraints on the lepton flavor
composition using the Fermi-LAT $\gamma$-ray measurements.
We find that the flavor composition is consistent with the Fermi-LAT data at
relatively low Galactic latitudes, while the fraction of the final state
$\tau^\pm$ is severely bounded.
\\[7mm]
Nucl.\ Phys.\ B\,(2020), in Press [arXiv:2004.10683 [astro-ph.HE]]
\end{abstract}


\newpage
\renewcommand{\thefootnote}{\arabic{footnote}}
\setcounter{footnote}{0}
\setcounter{page}{2}

\vspace*{10mm}
\tableofcontents

\setcounter{footnote}{0}
\renewcommand{\thefootnote}{\arabic{footnote}}

\baselineskip 18pt

\newpage
\section{Introduction}
\vspace*{1.5mm}
\label{sec:intro}
\label{sec:1}


High energy cosmic ray electrons and positrons (CRE)
can provide important information
for possible signals of dark matter (DM) annihilations or decays\,\cite{inDM}
and for the nearby galactic sources\,\cite{source}.
There have been very active experimental activities to measure the
cosmic ray electron/positron spectrum up to TeV energy scale,
including the ground-based and space-borne experiments
such as HESS\,\cite{HESS}, VERITAS\,\cite{VERITAS},
Fermi-LAT\,\cite{Fermi}, AMS-02\,\cite{AMS}, and CALET\,\cite{CALET}.
Measuring the high energy CRE spectrum opens up an
important window for the indirect detection of DM particles.

\vspace*{1mm}

Among these experimental activities, the first announcement of
the CRE energy spectrum measurement
from 25\,GeV up to 4.6\,TeV\,\cite{DAMPE2017}
by the Dark Matter Particle Explorer (DAMPE) detector\,\cite{DAMPE}
provided some intriguing results.
The DAMPE satellite was launched in December, 2015 and
is optimized for detecting cosmic $e^\pm$ events
and $\gamma$ rays up to about 10\,TeV energy.
The first CRE result of DAMPE collaboration was based on 530\,days of
data-taking\,\cite{DAMPE2017}.
The fitted CRE spectrum shows a spectral break around 0.9\,TeV and
is consistent with the HESS result\,\cite{HESS}. The DAMPE CRE data also
show a tentative peak-like event excess around $(1.3\!-\!1.5)$\,TeV,
which stimulated numerious studies on its possible
interpretations\,\cite{DAMPE-imply}-\cite{models},
ranging from the conventional astrophysical sources
(including pulsars and supernova remnants)
to models of DM annihilations or decays
into $e^+e^-$ events\,\cite{review}.

\vspace*{1mm}

In a previous short letter\,\cite{PLB},
we revealed a significant new hidden excess in the
energy region $(0.6\!-\!1.1)$\,TeV on the left-hand-side of the peak bin
$(1.3\!-\!1.5)$\,TeV, and we proposed that this new excess originates
from decays of the 1.5\,TeV $\mu^\pm$ ($\tau^\pm$)
events which are produced together with the 1.5\,TeV $e^\pm$ peak events.

\vspace*{1mm}

In this work, we perform a systematic study of the flavor structure of
the CRE spectra and study their implications for the DM indirect searches
including the DAMPE, Fermi-LAT and CALET experiments,
which fully go beyond \cite{PLB}.
We first make improved analysis by using the physical CRE backgrounds,
which is composed of the primary electrons produced by galactic supernova
remnants (SNR) and the secondary electrons from the collision between
primary nuclei and interstellar medium (ISM).
Then, we study annihilations of TeV scale DM particles into charged lepton
pairs ($e^+e^-,\,\mu^+\mu^-,\,\tau^+\tau^-$) in a nearby subhalo.
In addition to the possible peak-like structure measured by DAMPE,
we pay special attention to the possible non-resonant excess mentioned above.
We will show that the resultant CRE spectrum is consistent with the current
CRE data of DAMPE and Fermi-LAT, and also the AMS-02 and CALET data.
We demonstrate that including the muon decay channel not only
improves the fitting quality, but also gives important constraint on
the flavor structure of final state leptons from DM annihilations.
Finally, we take into account the photon radiation of the final state leptons,
and derive nontrivial constraints on the flavor compositions of the final
state leptons from the $\gamma$ ray measurements by the Fermi-LAT collaboration.

\vspace*{1mm}

This paper is organized as follows.
In Section\,\ref{sec:2}, we first revisit our previous study\,\cite{PLB},
in which we fit the DAMPE data with a double-broken power-law background and
with CRE signals from $\,\chi\chi\!\to\!e^+e^-\!,\mu^+\mu^-\!,\tau^+ \tau^-$ channels.
Then, we perform improved analysis with a physical CRE background
and the CRE signal spectrum from DM annihilations in both $\,e^+e^-$
and $\mu^+\mu^-(\tau^+ \tau^-)$ channels.
With these we derive constraints on the flavor composition of the
final state leptons from DM annihilations. We further estimate the increased
sensitivities by the projected future DAMPE running.
In Section\,\ref{sec:3}, we fit the Fermi-LAT, AMS-02 and CALET data,
and show the consistency with the $\mu^\pm$ ($\tau^\pm$) decay contribution
invoked for explaining the DAMPE data.
In Section\,\ref{sec:4}, we further use the $\gamma$-ray measurement of Fermi-LAT
to constrain the flavor composition of the final state leptons.
Finally, we conclude in Section\,\ref{sec:5}.

\vspace*{2mm}
\section{Improved Fit to DAMPE CRE Spectrum}
\label{sec:DAMPE}
\label{sec:2}
\vspace*{1mm}

In our previous short letter\,\cite{PLB},
we used an empirical broken power-law
formula to fit the background for analyzing the DAMPE data.
In this section, we will use the better
justified CRE physical background instead and perform a more realistic
analysis. In addition, we allow the DM mass as a free parameter
for the fit. Finally, we present an analysis of the projected
sensitivities for the future DAMPE running.
The systematic analyses of this section fully go beyond 
our previous short letter\,\cite{PLB}.

\vspace*{1mm}
\subsection{Fit with Broken Power-Law Background}
\label{sec:PLB}
\label{sec:2.1}
\vspace*{1mm}

\begin{figure}[t]
\centering
\vspace*{-3mm}
\includegraphics[height=8cm]{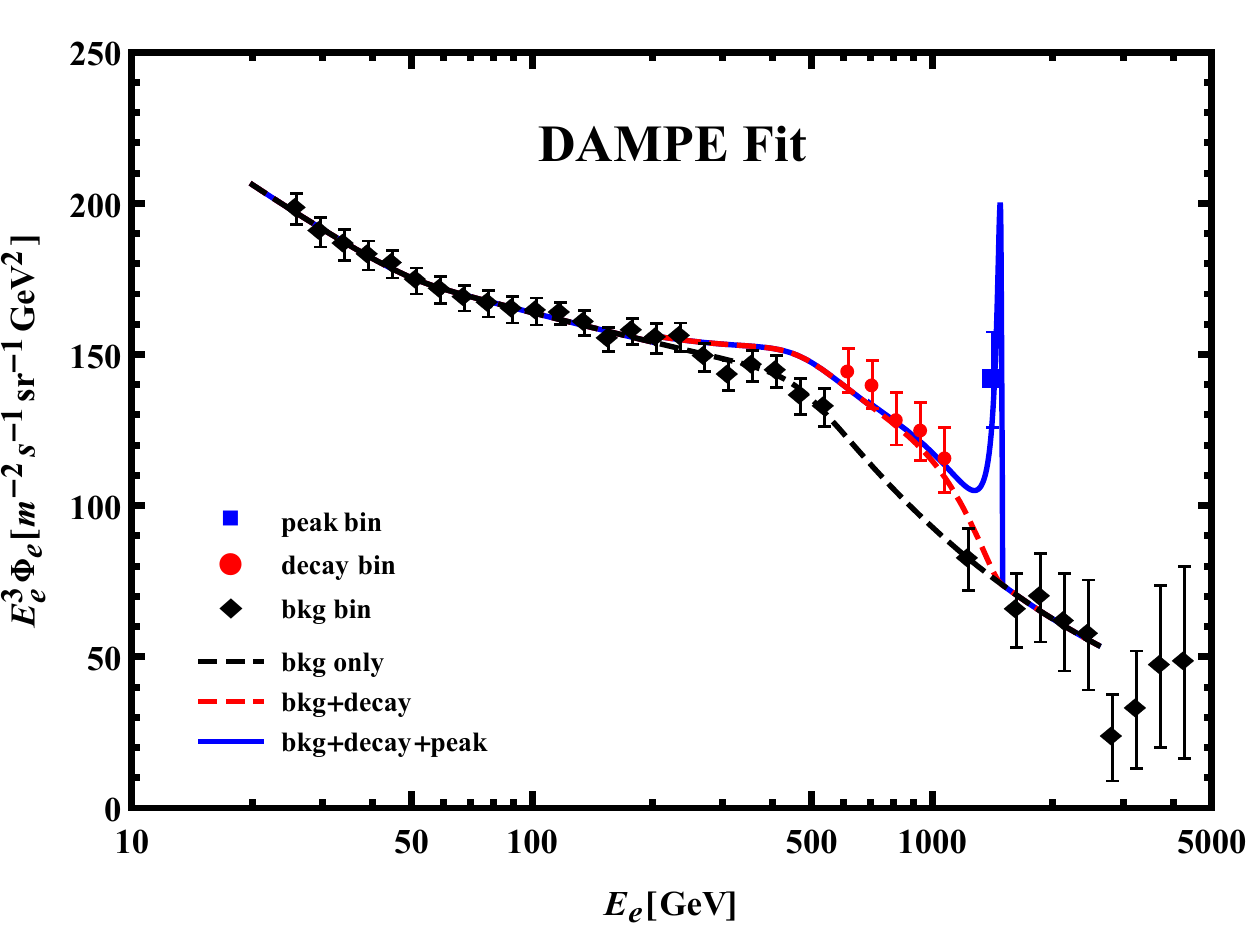}
\vspace*{-2mm}
\caption{\small
Fitting the CRE spectrum of DAMPE data. The fit of the background contribution is
shown by the black dashed curve; the fit further including the decay contribution
from 1.5\,TeV $\mu^\pm$ ($\tau^\pm$) composition is depicted by the red dashed curve;
and the fit further including 1.5\,TeV $e^\pm$ peak-like contribution is given by
the blue curve\,\cite{PLB}. The DAMPE data\,\cite{DAMPE2017} are shown by the
dots with $\pm 1\sigma$ error bars.
}
\label{fig:PLB}
\label{fig:1}
\vspace*{2mm}
\end{figure}

In this subsection, we first review what was done in Ref.\,\cite{PLB}, as
a comparison with our new analysis in the current study.
For clarity of our presentation, we replot in Fig.\,\ref{fig:1}
all the DAMPE data points with $\pm 1\sigma$ errors\,\cite{DAMPE2017}.
From this, we observed\,\cite{PLB} that the DAMPE data points
exhibit another rather intriguing structure
on the left-hand-side of the peak region $(1.3-1.5)$~TeV.
We found that the energy range of $(0.616\!-\!1.07)$~TeV
contains five consecutive data points (marked in red color), which
all lie above the background curve (as fitted from the other background
points marked in black color).
These five red data points are distinctive
and form a non-peak-like {\it new excess}
with $\gtrsim\!2\sigma$ significance.

\vspace*{1mm}

In Ref.\,\cite{PLB}, an empirical broken power-law formula
is adopted to describe the CRE backgrounds, without taking into
account their origin (including the sources and the propagation effect).
We first fit the background without including the five red data
points over $(0.616\!-\!1.07)$~TeV and the blue peak point at
$(1.3\!-\!1.5)$~TeV shown in Fig.\,\ref{fig:1}. The fit gives
a minimum $\chi^2$ per degree of freedom (d.o.f.),
$\,\chi^2\!/\text{d.o.f}\!=\!3.95/23 \!=\! 0.172$\,.
We show the background fit as the black dashed curve
in Fig.\,\ref{fig:1}.

\vspace*{1mm}

We then fit all the data points over the $25\,\text{GeV}\!-\!2.6\,$TeV region
by adding the DM annihilation contributions
$\,\chi\chi \!\to e^+e^-$\, and
$\,\chi\chi\!\to \mu^+ \mu^-(\tau^+ \tau^-) \!\to e^+e^-$\,
to the background.
For this analysis, we consider the DM particles with mass $\,M_\chi^{}\!=1.5\,$TeV
and suppose that they are accumulated in a nearby DM subhalo
whose density distribution is described by the generalized NFW profile\,\cite{NFW},
$\,\rho_\chi^{}(r) \equiv
\rho_s^{} (r/r_s^{})^{-\gamma}(1 \!+ r/r_s^{})^{\gamma-3}$,\,
with $\,\rho_s^{}\!=100\,\text{GeV}/\text{cm}^3$,\,
$r_s^{}\!=0.1$\,kpc, and $\gamma=0.5$\,.\,
We set the distance between the earth and the subhalo center as
$\,d_s^{}\!=0.2\,$kpc\,.\,
For the CRE spectra at source, the annihilation channel
$\,\chi \chi \!\to\! e^+ e^-$\, produces
$e^\pm$ flux with exactly 1.5\,TeV energy,
while the energy spectrum of $e^\pm$ from
$\,\chi\chi \!\to\! \mu^+ \mu^-(\tau^+ \tau^-) \!\to\! e^+ e^-$ channels
is given by
\beqa
\frac{1}{N_e^{}}\frac{\d N_e^{}}{\d E_e^{}}\,\simeq\,  \frac{4}{M_\chi^{}}\!
\(\!\frac{5}{12}-\frac{3E_e^2}{4M_\chi^2}+\frac{E_e^3}{3M_\chi^3}\!\)\!,
\eeqa
where the final states $e^\pm$ arise from the 3-body-decays
$\,\mu\!\to\!e\bar{\nu}_e^{}\nu_\mu^{}\,$ or
$\,\tau\!\to\!e\bar{\nu}_e^{}\nu_\tau^{}\,$.\,
We note that $\mu$ and $\tau$ share almost the same decay spectra
due to their energy
$\,E=\MX \!\gg m_{\mu}^{},m_{\tau}^{}$,\,
except for their different decay branching fractions
$\,\text{Br}[\mu\!\to\!e\bar{\nu}_e^{}\nu_\mu^{}]\!\simeq\! 100\%$\,
and
\,Br[$\tau\!\to\!e\bar{\nu}_e^{}\nu_\tau^{}]
\!\simeq\! 17.82\% \!\simeq\!1/5.6$ \cite{PDG}.
The lifetime of flying $\mu^\pm$ with energy $\,E_\mu^{}\!=\!1.5$\,TeV\,
is about $0.031$s
and could travel a distance about $9.36\!\times\!10^6$\,m.
The lifetime of $\,\tau^\pm$\, is shorter than $\,\mu^\pm$\,
by another 7 orders of magnitude.
Thus, the distances which $\mu^\pm$ and $\tau^\pm$ fluxes could travel
before their decays are negligible as compared to
$\,r_s^{}$ and $\,d_s^{}$\,.\,
So we can treat the initial $e^\pm$ fluxes from all channels
as produced at the source.
For the flux propagating from the source ${\bf x}_s^{}$
to a position ${\bf x}$\,,\, it can be formulated as
\beqa
\Phi_e^{}(E_e^{})
=
\int\!\! \d^3 x_s^{}\!\! \int \!\!\d E_s^{}\,
G({\bf x}, E_e^{}; {\bf x}_s^{}, E_s^{})\,
Q({\bf x}_s^{}, E_s^{}) \,,~~~
\eeqa
where $\,G({\bf x}, E_e^{}; {\bf x}_s^{}, E_s^{})$\, is the Green's
function and $\,Q\,$ is the source function. (The propagation part will
be further described in Section\,\ref{sec:2.3}.)
We fit the $e^\pm$ and $\mu^\pm$ ($\tau^\pm$) spectra together
for the DAMPE data.
This fit gives $\,\chi^2\!/\text{d.o.f}=16.5/27=0.609$\,,\,
which has a better quality than any other naive fits (with the peak
bin included).

\vspace*{1mm}

From the above fit, we also derive the thermally averaged cross sections
of the DM annihilations:
$\left\langle \sigma v \right\rangle_e
\!\!=\! 1.72\!\times\!10^{-26}\,\text{cm}^3/\text{s}$\,
and
$\left\langle \sigma v \right\rangle_\mu^{}\!
+0.178\left\langle \sigma v \right\rangle_\tau^{}
\!=\! 1.47\!\times\!10^{-25}\,\text{cm}^3/\text{s}$\,.\,
Their ratio (with $1\sigma$ bound) is given by
$\,y = y_\mu^{}\!+y_\tau^{}= 8.6^{+1.4}_{-2.5}$\,,\, where
$\,y_\mu^{}\!=\left< \sigma v \right>_\mu^{}\!/\!\left<\sigma v \right>_e$\,
and $\,y_\tau^{}\!=0.178\left< \sigma v \right>_\tau^{}\!/\!\left<\sigma v\right>_e$\,.\,
We further deduce the 90\% confidence limit (C.L.),
$\,y=2.6-10.8$\,.
For lepton portal DM models\,\cite{DAMPE-imply,LPDM},
we derive a non-trivial bound on the lepton-DM-mediator couplings
$\lambda_j^{}$ ($j=e,\mu,\tau$) \cite{PLB},
\beqa
\lambda_e^{} \!:\! \(\!\lambda_\mu^4
+\frac{1}{6}\lambda_\tau^4\!\)^{\!\!\frac{1}{4}}
=1\!:\!y^{\frac{1}{4}}\,,
\eeqa
with a fairly narrow range $\,y^{\frac{1}{4}}\simeq 1.3\!-\!1.8\,$ (90\%\,C.L.).
%

\vspace*{1mm}
\subsection{Physical Spectra for the CRE Backgrounds}	
\label{sec:bkg}
\label{sec:2.2}
\vspace*{1mm}

In the previous subsection, we discussed fitting the CRE backgrounds
with the empirical broken power-law formula as in Ref.\,\cite{PLB}.
For the current study, we will consider the physical components of the
CRE backgrounds. The CRE backgrounds consist of the primary electrons
from supernova remnants (SNR) and the secondary electrons/positrons
produced by inelastic hadronic interactions of cosmic ray nuclei
in the interstellar medium (ISM)\,\cite{review}.
They can be estimated numerically by the L{\scriptsize IKE}DM code\,\cite{LikeDM}.

\vspace*{1mm}

The primary electrons are considered to mainly arise from the SNR.
Their injection spectrum can be formulated as\,\cite{LikeDM},
\begin{eqnarray}
\label{eq:bkg}
  \frac{\d N}{\d E_s^{}} \,=\,
  A_e^{} \left(\frac{E_s}{1\text{GeV}}\right)^{\!\!\!-\gamma_2^{}}\!
\left[ 1 + \left(\!\frac{E_s^{}}{\,E_\text{br2}^{}} \right)^{\!\!\!2\,}
  \right]^{\!\!\frac{\,\gamma_2^{} - \gamma_3^{}\,}{2}}
  \!\!\exp\!\left(\!\!-\frac{E_s^{}}{E_\text{cut}^{}\,}\!\right) \!,
\end{eqnarray}
for $\,E_s^{}\!\gtrsim\! 20\,\mbox{GeV}$. To estimate the primary
electron background spectrum after propagation, we first approximate
the injection spectrum with a series of Gaussian kernels,
\begin{eqnarray}
\frac{dN}{dE_s} \,=\,
\sum_{j=1}^{n} \frac{C_j^{}}{\sqrt{2\pi\,}\sigma_j^{}\,}
\exp\!\left[-\frac{(E_s^{}\!-\!E_j^{})^2}{2\sigma_j^2} \right]\!,
\end{eqnarray}
where $\,E_j^{}=0.01\!\times\!10^{(j-1)/7}$\,GeV\,
and $\,\sigma_j^{}\!=0.15E_j^{}$\, for $n=50$ \cite{LikeDM}.
The post-propagation flux for each Gaussian kernel, $\Phi_i^{}(E_e^{})$,
is included in the L{\scriptsize IKE}DM code.
For the present analysis, we will adopt its third set of
the propagation parameters\footnote{{Recently, the HAWC
observations of extended $\gamma$-ray halos around pulsars indicate
that the particle diffusion in the Milky Way is likely inhomogeneous
\cite{pulsar}. However, the filling volume of such inefficient
diffusion regions should be small to be consistent with the cosmic
ray observations. Furthermore, the very high energy CREs observed
by HESS also suggest that the diffusion in the local region of the
Earth should be fast\,\cite{Hooper:2017tkg}. Hence, we use the
conventional homogeneous diffusion model for this study.}},
in which the diffusion coefficient is parametrized
as $\,D(x,E) = D_0^{} (E/4\text{GeV})^\delta$ with
$\,D_0^{} = 7.1\!\times\! 10^{28}\,\text{cm}^2/\text{s}$\,
and $\,\delta = 0.33$\,\cite{LikeDM}.
Thus, we obtain the total primary electron flux as a linear combination
\begin{eqnarray}
\Phi(E_e^{}) \,=\,
\sum_{j=1}^{n} C_j^{} \Phi_j^{}(E_e^{})\,.
\end{eqnarray}

The secondary $e^+$ flux is computed according to the cosmic ray
proton and Helium nuclei interactions with the ISM during their propagation.
The secondary $e^-$ shares almost the same spectrum as the secondary $e^+$,
but has an additional suppression factor of 0.6 \cite{LikeDM}.
This spectrum is also provided by the L{\scriptsize IKE}DM code.
Since the solar modulation effect is only significant in the low energy
region $E<10$\,GeV, it can be neglected in our fit.

\vspace*{1mm}
\subsection{\,CRE Spectra from DM Annihilations}	
\label{sec:signal}
\label{sec:2.3}
\vspace*{1mm}

\begin{figure*}[t]
\centering
\vspace*{-1mm}
\includegraphics[height=6.0cm]{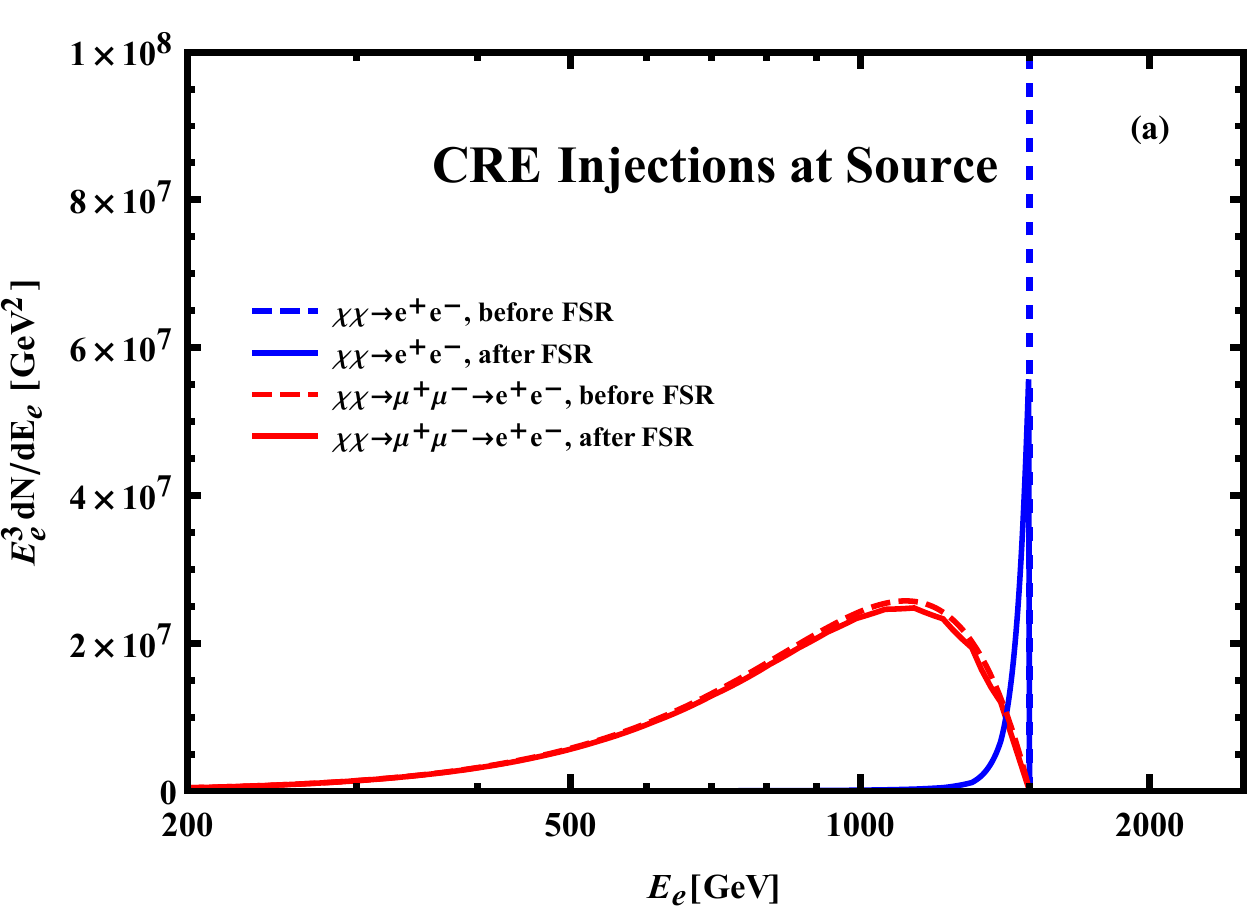}
\hspace*{2mm}
\includegraphics[height=6.0cm]{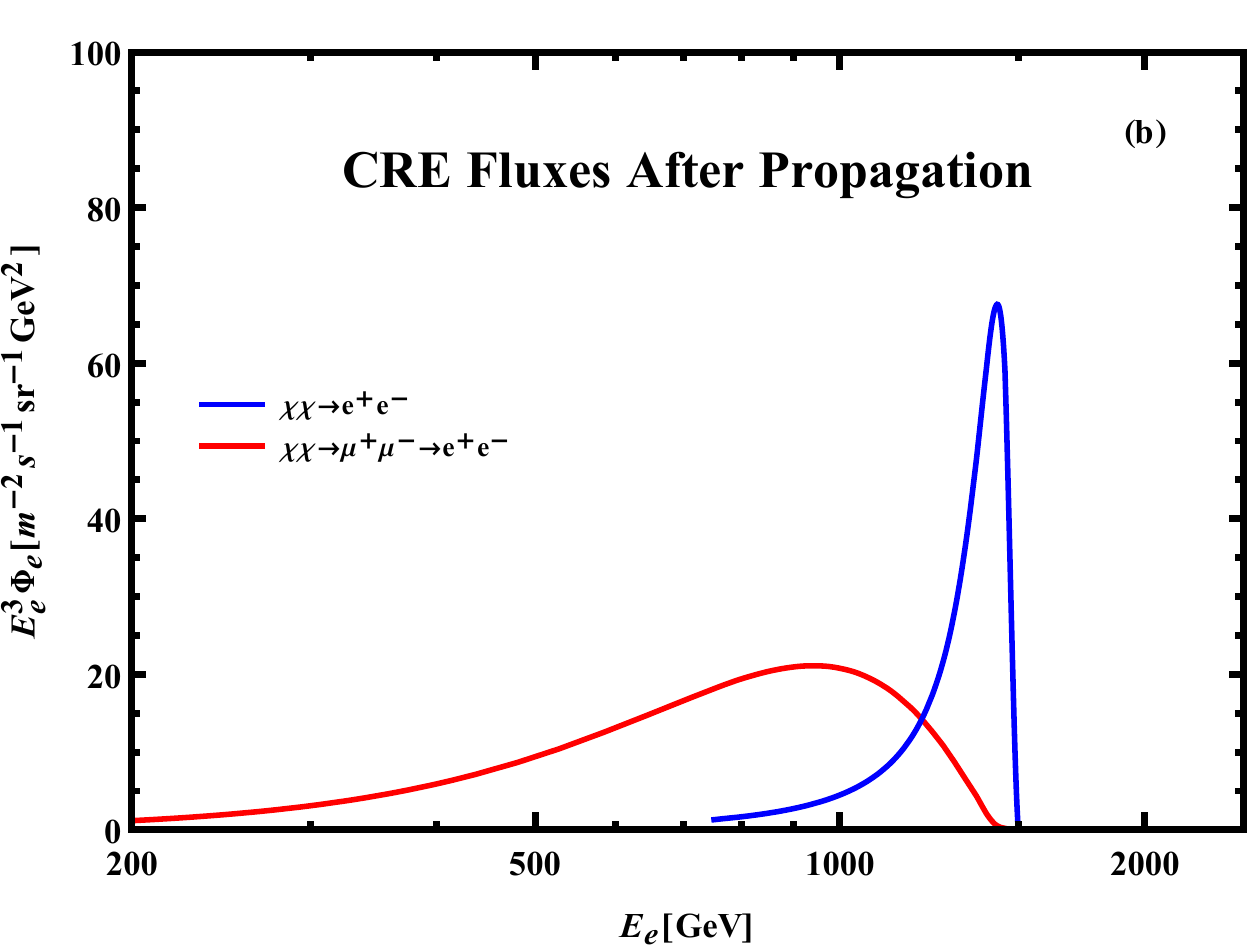}
\vspace*{-2mm}
\caption{\small
CRE energy spectrum at source and the fluxes after their propagation to the earth.
Plot\,(a) presents the injection spectra of the CRE from DM annihilations.
The blue curves show the $e^\pm$ channel and the red curves present
the $\mu^\pm$ decay channel.
The spectra after including the FSR are given by solid curves, while those
without FSR are shown by dashed curves. For clarity, the spectra of $\mu^\pm$
decay channel is rescaled by an enhanecment factor of 50.
Plot\,(b) presents the CRE fluxes after the propagation to the earth.
For illustration, the thermally averaged annihilation cross section of each channel
is chosen as $\left<\sigma v \right>_e^{}\!=2.0\!\times\!10^{-26}\,\text{cm}^3\!/s$
and $\left<\sigma v \right>_\mu^{}\!=1.5\!\times\!10^{-25}\,\text{cm}^3\!/s$,\,
while the DM mass is set as $M_\chi^{}\!=1.5\,$TeV for both plots.
}
\label{fig:propagation}
\label{fig:2}
\end{figure*}

We use the PPPC4DMID\,\cite{PPPC}\cite{Flux} package
to calculate the electron spectra of the DM annihilation processes
$\chi\chi\rightarrow e^+e^-$ and
$\chi\chi\rightarrow\mu^+\mu^-(\tau^+\tau^-)\rightarrow e^+e^-$.\,
The final state radiation (FSR) in the DM annihilation process will soften the CRE spectra.
In Fig.\,\ref{fig:2}(a), we show the injection spectra before and after FSR
as dashed and solid curves, respectively.
For clarity of presentation, we also rescaled the spectra of $\mu^\pm$
decay channel by an extra enhanecment factor of 50.
Here we do not distinguish the spectra of $\mu^\pm$ and $\tau^\pm$ channels,
since their shapes are rather similar except that the $\tau^\pm$ channel
is suppressed by the much lower branching fraction of the 3-body-decays of $\tau^\pm$.

\vspace*{1mm}

Next, we analyze the propagation of the $e^\pm$ flux from the nearby DM subhalo.
When traveling across the interstellar space, CREs experience diffusion and
energy loss, as described by the following propagation equation,
\beqa
\frac{\,\partial N_e^{}}{\partial t}
- \frac{\,\partial [b(x,E) N_e]\,}{\partial E}
- \nabla (D(x,E) \nabla N_e^{})
\,=\,  Q \,,
\label{eq:diffusion}
\eeqa
where $\,N_e^{} (E_e, t, {\bf x})$\, is the number density distribution as a function
of the $e^\pm$ energy $E_e$ and the spacetime coordinates $(t, {\bf x})$.

\vspace*{1mm}

The energy loss, $\,b(E) \equiv - \d E/\d t$\,,\,
is defined as\,\cite{PPPC}:
\begin{eqnarray}
b(x,E) \,=\, \frac{\,4\sigma_T^{}\,}{\,3m_e^2\,}
E^2\!\left[u_B^{}(x)+\sum_i u_\gamma^{(i)}(x,E) \right]\!,
\end{eqnarray}
where $\,u_B^{}=\frac{1}{2}B^2\,$  stands for the energy density contribution
from the galactic magnetic fields,
and $u_\gamma^{(i)}$ is the energy density of photons from the CMB,
starlight and dust-diffused infrared light.
In the above equation, $\,\sigma_T^{}={8\pi\alpha_\text{em}^2}/(3m_e^2)$\,
is the Thompson cross section.
Since the size of the DM subhalo and its distance to earth are fairly small as
compared to the galaxy size, we can ignore the $x$ dependence of $b$\,.\,
Furthermore, at the location of solar system, which is about 8\,kpc away from
our galaxy center,
the galactic magnetic field is $\,B\!\sim\!10^{-6}$\,G,
and hence $u_B^{}$ is negligible.
Thus, the energy-loss term can be rewriten as
\begin{eqnarray}
b(x,E) =
\frac{~4\sigma_T^{}\sum_i^{}\!u_\gamma^{(i)}~}{3m_e^2} E^2
\,=\, b_0^{} \!\(\!\!\frac{E}{\,\text{GeV}\,}\!\!\)^{\!\!2},
\end{eqnarray}
with $\,b_0^{} = 10^{-16}\,\text{GeV}\!/\text{s}$\,.\,

\vspace*{1mm}

For the diffusion coefficient $D(x,E)$,
we use the same parametrization as in Section\,\ref{sec:2.2}.
After conversion of units, we have
\begin{eqnarray}
D(E) = 149 \!\left(\frac{E}{\text{GeV}}\right)^{\!\!0.33}
\text{pc}^2\!/\text{kyr}\,.
\end{eqnarray}

\vspace*{1mm}

The right-hand-side of Eq.\eqref{eq:diffusion} is the $e^\pm$ source function,
\beqa
Q({\bf x}, E_e^{}) \,=\,
C \frac{\,\left\langle\sigma v \right\rangle\!\rho_\chi^2({\bf x})\,}{M_{\chi}^2}
\frac{\d N}{\,\d E_e^{}\,},\,
\eeqa
where $\,\rho_\chi^{}({\bf x})\,$ is the DM density distribution,
$M_{\chi}^{}$ is the mass of the DM particle,
$\,\left\langle \sigma v \right\rangle\,$
is the thermally averaged annihilation cross section,
and $\,\d N/\d E_e^{}\,$ the $e^\pm$ is the energy spectrum from the DM annihilations.
The coefficient $\,C\!=\!\frac{1}{4}$\, for Dirac fermion or complex scalar DM, and $\,C\!=\!\frac{1}{2}\,$ for Majorana fermion or real scalar DM\,\cite{models}.
For the current illustration, we will set
$\,C\!=\!\frac{1}{4}$\, hereafter.
In this work, we use the generalized NFW density profile
as in Ref.\,\cite{PLB},
$\,\rho_\chi^{}(r) \equiv
\rho_s^{} (r/r_s^{})^{-\gamma}(1 \!+\! r/r_s^{})^{\gamma-3}$,\,
with $\,\rho_s^{}=100\,\text{GeV}/\text{cm}^3$,
$r_s^{}=0.1$\,kpc and $\gamma=0.5$.\,
We set the distance between the earth and the center of the subhalo as
$\,d_s^{}\!=0.2\,$kpc.

\vspace*{1mm}

The propagation function (\ref{eq:diffusion})
can be solved with the Green's function\,\cite{diffusion},
\beqa
G({\bf x}, E; {\bf x}_s^{}, E_s^{})
\,=\,
\frac{~\exp\!\left[ - |{\bf x} - {\bf x}_s^{}|^2/\lambda^2 \right]~}
{b(E) (\pi \lambda^2)^{3/2}} \,,
\eeqa
where $E_s^{}$ is the $e^\pm$ energy at source, and $\,E\,$ is the
corresponding energy after propagation. The propagation scale
$\lambda\,$ is given by
\beqa
\lambda^2 \,=\, 4 \!\int^{E_s}_E\!\! \d E' \frac{\,D(E')\,}{b(E')}.
\eeqa
Then, the solution of Eq.(\ref{eq:diffusion}) can be expressed as
\beqa
N_e^{}({\bf x},E_e^{})
=
\int\!\! \d^3 {\bf x}_s^{}\!\! \int \!\!\d E_s^{}
G({\bf x}, E_e^{}; {\bf x}_s^{}, E_s^{})
Q({\bf x}_s^{}, E_s^{}) \,.~~~
\eeqa
Finally, the CRE flux $\Phi_e^{}$ is related to
the density distribution $N_e^{}$ by
\begin{eqnarray}
\Phi_e^{}({\bf x},E_e^{}) \,\equiv\, \frac{1}{4\pi} N_e^{}
({\bf x},E_e^{}) v(E_e^{})\,,
\end{eqnarray}
where $\,v(E_e^{})$\, is the velocity of electron with energy $E_e^{}$.
Since $\,E_e^{} \gg m_e^{}$ holds for the DAMPE data, we have $v\simeq c$\,.\,
In Fig.\,\ref{fig:2}(b),
we plot the CRE flux from all channels after $e^\pm$ signals propagate
to the earth.
For illustration, we chose the thermally averaged annihilation cross section
of each channel to be
$\left<\sigma v \right>_e^{}\!=2.0\!\times\!10^{-26}\text{cm}^3\!/{\rm s}$\,
and $\,\left<\sigma v \right>_\mu^{}\!=1.5\!\times\!10^{-25}\text{cm}^3\!/{\rm s}$,
and set the DM mass $M_\chi^{}\!=1.5\,$TeV.

\vspace*{2mm}
\subsection{\,Fitting Event Excesses in CRE Spectrum}	
\label{sec:fit}
\label{sec:2.4}
\vspace*{1mm}

The total $e^\pm$ flux contains three parts, i.e.,
$\Phi_e^{} \equiv \Phi_{\rm bkg}^{} \!+\! \Phi_{\rm decay}^{} \!+\! \Phi_{\rm peak}^{}$,
where the flux $\Phi_{\rm bkg}^{}$ denotes the CRE backgrounds,
$\Phi_{\rm decay}^{}$ arises from the decay contributions of the
$\mu^\pm$\,($\tau^\pm$) as produced from the DM annihilations, and
$\Phi_{\rm peak}^{}$ is the contribution of $e^\pm$
as produced by the DM annihilations.
We fit the DAMPE data points by minimizing the $\chi^2$ function,
\begin{eqnarray}
\label{eq:Dchi2}
\chi^2 \,=\, \sum_j\! \left[\frac{1}{\,\sigma_j^{}\,}\!
\left(\!\Phi_j^{}- \!\int_{E_j^l}^{E_j^u}\!\!\!\d E\,
\frac{\Phi_e^{}(E)}{\,E_j^u\!-\!E_j^\ell\,}\right)  \right]^{\!2}\!,
\end{eqnarray}
where $\Phi_j^{}$, $\sigma_j^{}$, $E_j^\ell$ and $E_j^u$ stand for
the detected CRE flux, the error, the lower and upper energy bounds
of the $j$th bin, respectively.

\vspace*{1mm}

\begin{figure*}[t]
\centering
\vspace*{-8mm}
\hspace*{-4mm}
\includegraphics[height=6.3cm]{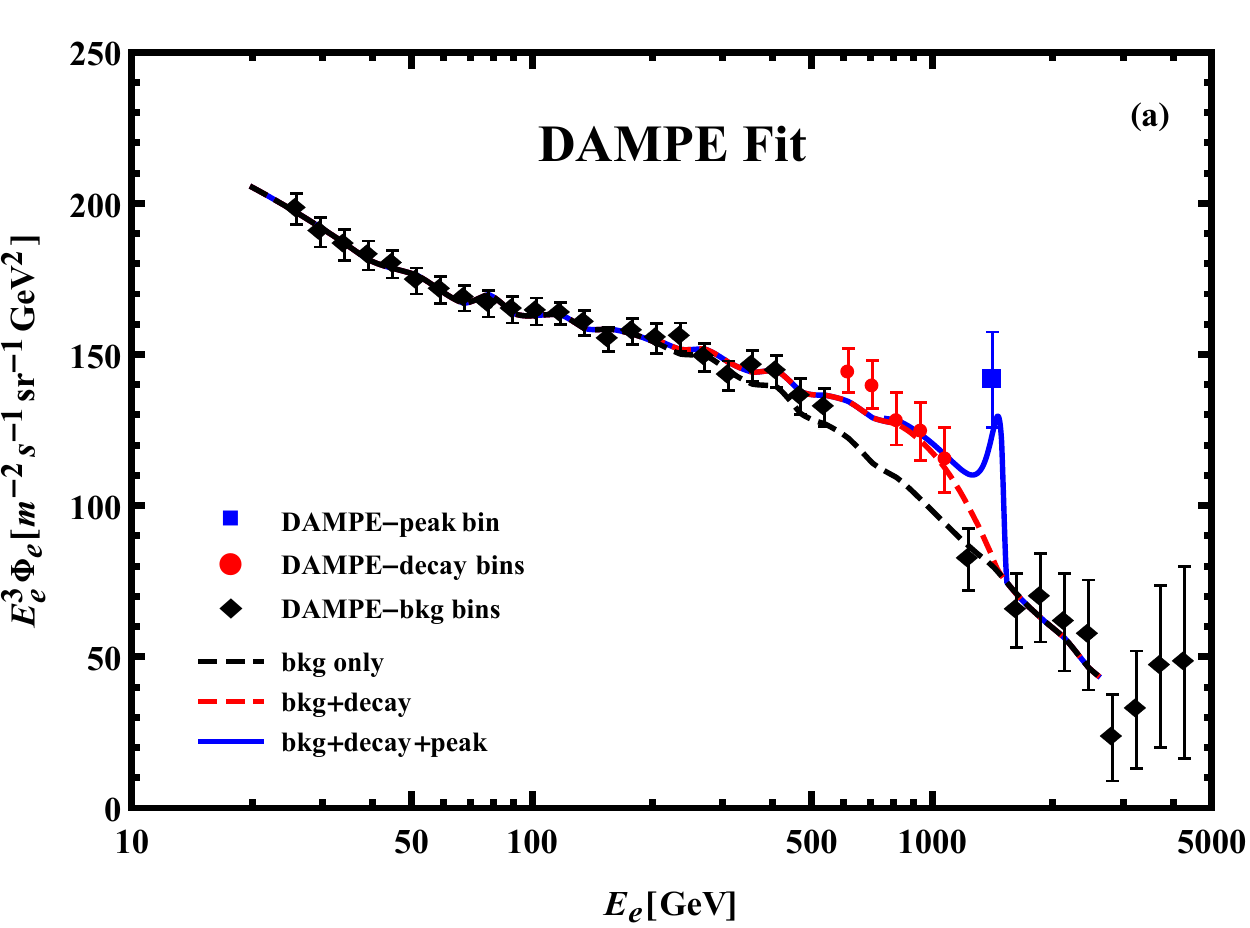}
\hspace*{1.5mm}
\includegraphics[height=6.0cm]{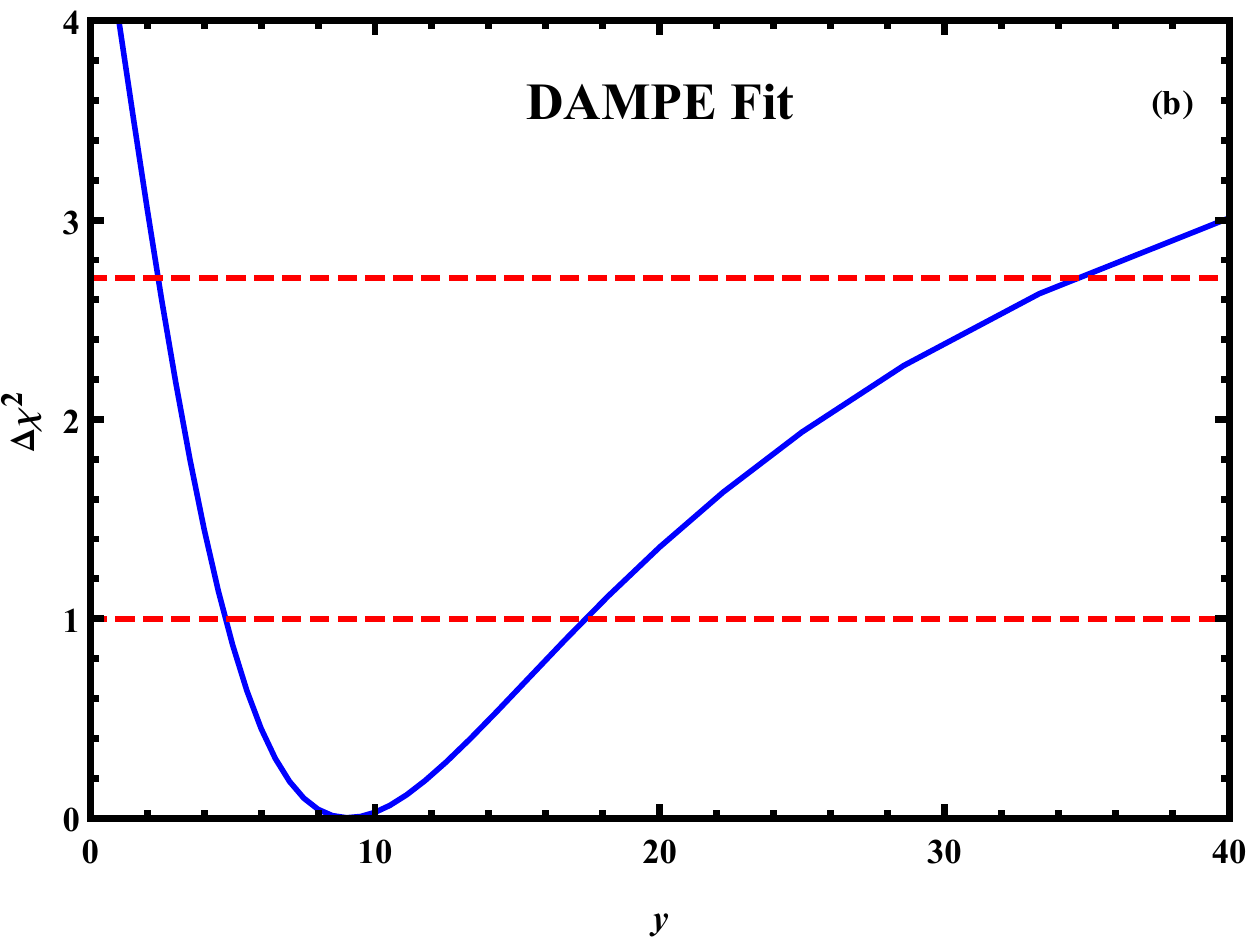}
\\[2mm]
\hspace*{-1.5mm}
\includegraphics[height=5.9cm]{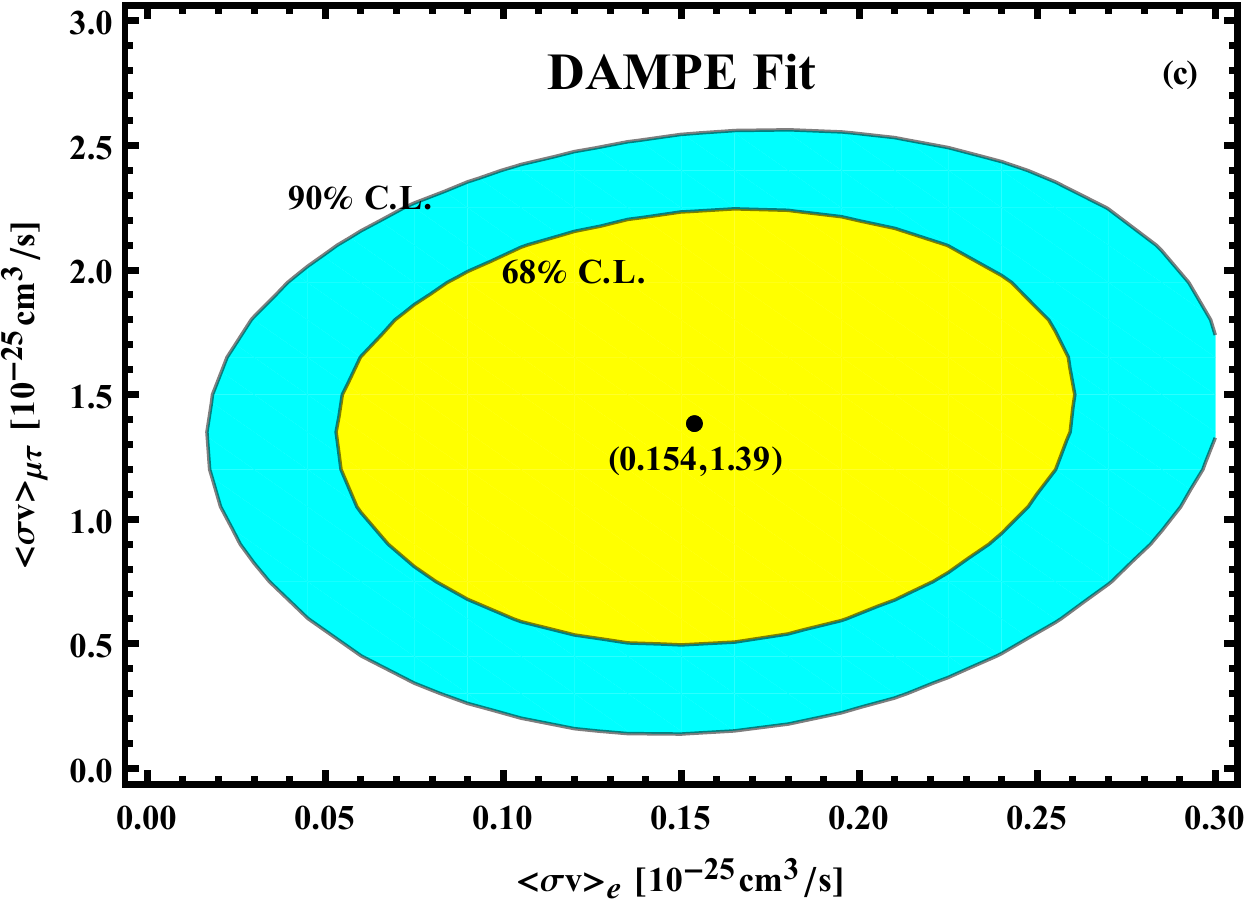}
\hspace*{0mm}
\includegraphics[height=6.0cm]{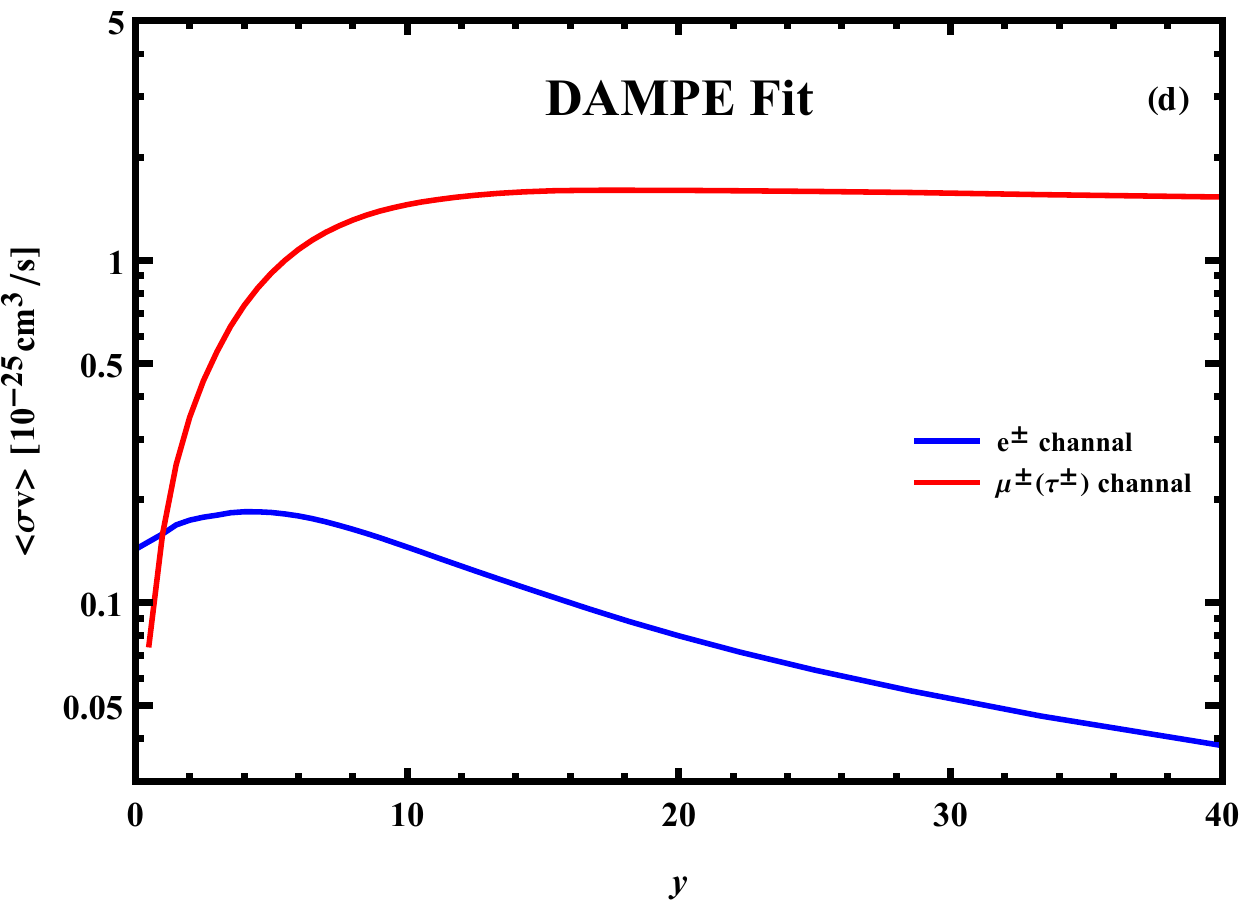}
\vspace*{-4mm}
\caption{\small
Plot\,(a): Improved fit for the DAMPE CRE spectrum,
with the central value of the flavor ratio $y=9.0$\,.\,
Plot\,(b): $\Delta\chi^2$ as a function of the flavor ratio $y$\,.
The allowed ranges of the flavor ratio are
$\,y=9.0^{+8.4}_{-4.2}$ ($1\sigma$) and
$\,2.4\!<\!y\!<\!34.7$ (90\%\,C.L.).
Plot\,(c): $\Delta\chi^2$ contours for the DM annihilation cross sections
of $e^\pm$ versus $\mu^\pm$\,($\tau^\pm$) channels.
The $\mu^\pm$\,($\tau^\pm$) decay contribution is nonzero at 90\%\,C.L.
Plot\,(d): The cross sections
$\left\langle \sigma v \right\rangle_e^{}\!$ and $\left\langle \sigma v \right\rangle_{\mu\tau}\!$
as functions of the flavor ratio $y$\,.\,
}
\label{fig:DAMPE-new}
\label{fig:3}
\vspace*{3mm}
\end{figure*}

Our best-fit result gives the DM mass $\,M_\chi^{}\!=\!1.54\,$TeV,\,
and the thermally averaged annihilation cross sections:
\beqs
\beqa
\left\langle \sigma v \right\rangle_e^{}
 \!&=&\!\! 1.54\!\times\!10^{-26}\text{cm}^3/\text{s}\,,
\\[1.5mm]
\left\langle \sigma v \right\rangle_{\mu\tau}^{}
 \!\!&=&\!\!
\left< \sigma v \right>_\mu^{} + \,0.178\left< \sigma v \right>_\tau^{}
 = 1.39\!\times\!10^{-25}\text{cm}^3/\text{s}\,.\,
\eeqa
\eeqs
The ratio of cross sections between the final state $\mu^\pm$ ($\tau^\pm$)
and the final state $\,e^\pm$\, is given by
$\,y = y_\mu^{}\!+y_\tau^{} \simeq 9.0$\,,\, with
$\,y_\mu^{}\!=\left< \sigma v \right>_\mu^{}\!/\!\left<\sigma v \right>_e$\,
and $\,y_\tau^{}\!=0.178\left< \sigma v \right>_\tau^{}\!/\!\left<\sigma v\right>_e$\,.\,
For the CRE background spectrum \eqref{eq:bkg},
this fit gives the following parameters:
\beqa
A_e^{}\!=0.130\,\text{GeV}^{-1},~~~
\gamma_2^{}=2.80,~~~
\gamma_3^{}=2.29,~~~
E_\text{br2}^{}\!=46.3\,\text{GeV},~~~
E_\text{cut}^{}\!=2.56\,\text{TeV}.\,
\hspace*{8mm}
\eeqa
The fitting quality is given by
$\chi^2/\text{d.o.f.}\!=\!19.6/26=0.756$\,,\,
which is fairly good.
We note that the largest deviation between the data and our model comes
from the $(1.1\!-\!1.3)$\,TeV bin in Fig.\,\ref{fig:3}(a),
i.e., the black bin lying between the red bins and the blue peak bin,
which is $2.8\sigma$ below the best fit value in the blue curve.
This contributes about 40\% of the total $\chi^2$ value.

\vspace*{1mm}

We present our new fit of the CRE spectrum including the $\mu^\pm$\,($\tau^\pm$)
decay contribution (red curve)
and the peak signal (blue curve) in Fig.\,\ref{fig:3}(a).
Impressively, this demonstrates that including the $\mu^\pm$\,($\tau^\pm$)
decay events can fully explain this non-peak-like new excess
over the $(0.6-1.1)$\,TeV energy region.
Furthermore, fitting the new excess consequently gives rise to 
the spectral break around 0.9\,TeV,
while leaving the background curve relatively flat.
This provides a DM interpretation of the spectral structure,
besides other astrophysical sources such as pulsar wind nebulae (PWN) 
or supernova remnants (SNR)\,\cite{break}.

\vspace*{1mm}

We can derive the constraints on the flavor ratio $y$ through marginalization.
We scan the value of $\,y\,$ in a wide range,
and compute the $\Delta\chi^2$ as a function of $\,y\,$.
This is plotted as the blue curve in Fig.\,\ref{fig:3}(b).
We find that the flavor ratio with $1\sigma$ bounds is
$\,y\!=9.0^{+8.4}_{-4.2}$\,,\, and its 90\% confidence limits are
$\,2.4\!<\!y\!<\!34.7$\,.\footnote{%
	This improved analysis differs from our previous fit\,\cite{PLB}, where
	we made a separate background fit, and then with the background
	parameter $E_{\text{br2}}^{}$ varying within its 90\% limits
	we further fitted the signal contributions.
	In the current analysis, we fit the background parameters and
	signal parameters {simultaneously}, hence it gives weaker
	but more reliable limits.}\,
If we set $y=0$\,,\, the best fit has its $\chi^2$ value increase to
$\,\chi^2\!/\text{d.o.f.}\!=24.7/27=0.914$\,.\,
For this case, the minimal $\chi^2$ increases by $\Delta\chi^2=5.02$,\,
in which the 5 non-resonant red bins contribute $\Delta\chi^2\!=\!4.68$\,.
This shows that including the $\mu^\pm$\,($\tau^\pm$) decay contribution
($y>0$) for fitting the 5 red bins does play an important role
to improve the fitting quality.

\vspace*{1mm}

In Fig.\,\ref{fig:3}(c), we marginalize the two cross sections,
$\left<\sigma v\right>_e^{}$ and
$\left<\sigma v\right>_{\mu\tau}^{}\!\!=\!
 \left<\sigma v\right>_\mu^{}\!\!+0.178\left<\sigma v\right>_\tau^{}$,\,
which determines the final state $e^\pm$ events from the 3-body decays of
the final states $\mu^\pm$ and $\tau^\pm$.\,
We show their 68\% and 90\% confidence limits as the yellow and light blue contours,
respectively.
Our fit demonstrates that a nonzero $\mu^\pm$\,($\tau^\pm$)
decay contribution is required at 90\%\,C.L.

\vspace*{1mm}

\begin{table*}[t]
	\centering
	\begin{tabular}{c||c|c|c}
		\hline\hline
		&&& \\[-3.5mm]
		Varying\,Peak\,Signals &
		$\left\langle \sigma v \right\rangle_e$ ($10^{-25}\text{cm}^3\!/\text{s}$)
		& $\left\langle \sigma v \right\rangle_{\mu\tau}^{}$ ($10^{-25}\text{cm}^3\!/\text{s}$)
		& Flavor Ratio $y$
		\\[1mm]
		\hline\hline
		&&& \\[-3.8mm]
		original\,value~~~~ & 0.154 & 1.39 & 9.0
		\\[0.7mm]
		\hline
		&&& \\[-3.8mm]
		$+1\sigma$ upper\,value & 0.215 & 1.30 & 6.1
		\\[0.7mm]
		\hline
		&&& \\[-3.8mm]
		$-1\sigma$ lower\,value~ & 0.094 & 1.47 & 15.7~
		\\[0.7mm]
		\hline\hline
	\end{tabular}
	\caption{\small
		Fluctuation of the peak signals versus the contribution of
		$\mu^\pm$ ($\tau^\pm$) channel.
		When the number of peak signal events fluctuates within $\pm 1\sigma$ range,
		the contribution of the $e^\pm$ channel varies accordingly (2nd column), but
		the contribution of $\mu^\pm$ ($\tau^\pm$)
		channel (3rd column) only changes by less than 6\%
		and remains nearly the same in the fit. The last column shows that the corresponding
		change of the flavor ratio $y$\, is mainly due to the fluctuation of peak signals.
	}
\label{tab:1}
\end{table*}

We also note that in Fig.\,\ref{fig:3}(c),
the shapes of the $\Delta\chi^2$ contours
are close to ellipses with two axes nearly parallel to
the horizontal/vertical directions,
which means that the statistical correlation
between the cross sections is quite weak.
This fact means that these two parameters may be fitted independently.
We illustrate this point in Table\,\ref{tab:1}.
Here we fluctuate the number of peak signal events by $\pm 1\sigma$
and find that although the contribution of $e^\pm$ channel varies
accordingly (2nd column),
the contribution of $\mu^\pm$ ($\tau^\pm$) channel only changes by
less than 6\% (3rd column) and remains nearly the same in the fit.
Then, we note that the flavor ratio $\,y\,$ (4th column)
has significant changes mainly due to the fluctuation of peak signals.
It shows that even if the annihilation rate to the $e^\pm$ channel
and hence the value of the flavor ratio $\,y\,$ may change significantly,
the contribution from the non-resonant $\mu^\pm$ ($\tau^\pm$) channel
is largely independent of the event number in the peak bin and fairly robust.
Finally, we note that the small change of the cross section
$\left<\sigma v \right>_{\mu\tau}^{}
[\,=\!\left< \sigma v \right>_\mu^{} \!+ 0.178\left< \sigma v \right>_\tau^{}]$\,
in the above fit mainly comes from the tension
in the non-resonant bin of $(1.1\!-\!1.3)$\,TeV (which is on the left-hand-side of the
peak bin). Namely, in Fig.\,\ref{fig:3}(a) we see that this $(1.1\!-\!1.3)$\,TeV bin
is significantly below the blue curve;
so reducing events in the peak bin will slightly soften the tension with this bin
and thus allow more $\mu^\pm$ ($\tau^\pm$) contribution.
From the above analysis, we find that
{\it the evidence of the non-peak excess can stand by itself, and is fairly
independent of the peak excess.}

\vspace*{1mm}

When making the $\chi^2$ fit in Fig.\,\ref{fig:3}(b) for each given $y$ value,
we also obtain the fit values for other free parameters as functions of $y$,
including the two cross sections.
We explicitly plot the two fitted cross sections in Fig.\,\ref{fig:3}(d).
From these curves, we observe that for
$\,y \!=\! y_\mu^{}\!\!+y_\tau^{}\!\gtrsim 9$
the $\mu^\pm$\,($\tau^\pm$) decay contribution remains fairly flat,
and the cross section $\,\left< \sigma v \right>_e^{}$ drops roughly as $\,y^{-1}$\,.
This means that for larger values of $y$\,,
the contribution from $\mu^\pm$\,($\tau^\pm$) decays is already maximized,
so the increase of $\,y\,$ is mainly caused by
the decrease of the $e^\pm$ cross section $\,\left<\sigma v \right>_e^{}$\,.
The fact that the contribution of $e^\pm$ channel gets suppressed
for large $\,y\,$ is important for our $\gamma$ ray analysis,
which will be discussed in Section\,\ref{sec:4}.

%
\vspace*{1mm}
\subsection{\,Origin of Flavor Composition of CRE Spectrum}	
\label{sec:ratio}
\label{sec:2.5}
\vspace*{1mm}

From the analysis in Section\,\ref{sec:2.4}, we find
that the original lepton final state produced at a nearby source
should have a flavor composition ratio,
$N_e^{} \!:\! (N_\mu^{}\!+0.178N_\tau^{}) = 1\!:\!y$\,,\,
with the constrained range $\,2.4<y<34.7\,$ at 90\%\,C.L.
(Our analysis here used the current improved fit of $\,y\,$ which
differs from our previous analysis\,\cite{PLB} and is more reliable.)
We note that in the above ratio,
the $\tau$ component could only play a minor role
due to the suppression by its small decay branching fraction
$\,\text{Br}[\tau\!\to\!e\,\bar{\nu}_e^{}\nu_\tau^{}]\!\simeq\!17.82\%\,$.\,
The simplest realization of this flavor composition condition is
$\,N_e^{} \!:\! N_\mu^{} \!:\! N_\tau^{} \approx 1\!:\! y\!:\!0$\,,\,
where all the non-peak excesses are given by the decay contribution of muon flavor.

\vspace*{1mm}

The above flavor composition condition will place important
constraint on the lepton-related DM model building.
For instance, we can consider the typical lepton portal
DM models\,\cite{DAMPE-imply,LPDM}, where
the DM particle is either a fermion or scalar.
For the case of fermionic DM, we consider
a neutral singlet Dirac fermion $\chi$ as the DM,
which interacts with a scalar mediator $S$ and the right-handed charged lepton
$\ell_{Rj}^{}$ as follows:
\beqa
\label{eq:DM-fermion}
{\cal L}_{\chi}^{}\supset \lambda_j^{}S_j^{}\,\overline{\chi^{}}\ell_{j}^{}
+\text{h.c.}\,,
\eeqa
where $\,\ell_j^{}=e,\mu,\tau$.\,
For the case of scalar DM, consider a neutral complex singlet scalar $X$
as the DM particle and a Dirac fermion $\psi$ as the mediator
(which has the same electric charge and lepton number as the charged leptons).
Thus, the relevant interaction Lagrangian takes the following form:
\beqa
\label{eq:DM-scalar}
{\cal L}_{X}^{}\supset
\la_j^{}X\overline{\psi_{j}^{}}\ell_{j}^{}+\text{h.c.}
\eeqa

The above interactions will realize the DM annihilation process
$\chi\bar{\chi}\to \ell_j^{}\ell_j^{}$
or $XX\to \ell_j^{}\ell_j^{}$
via the $t$-channel exchange of $S_j^{}$ or $\psi_{Lj}^{}$.\,
We note that the cross section of DM annihilations is proportional to
$\,\lambda_j^4\,$.\,
For the above simplest realization of the flavor composition condition,
we can deduce
$\,N_e \!:\! N_\mu \!:\! N_\tau
= \lambda_e^4\!:\! \lambda_\mu^4 \!:\! \lambda_\tau^4
= 1\!:\! y\!:\!0$\,,\, where the flavor ratio $\,y$\, is constrained within
{$\,2.4 < y < 34.7\,$} at 90\%\,C.L.
Hence, this results in a simple coupling relation,
$\,\lambda_e^{}\!:\! \lambda_\mu^{} \!:\! \lambda_\tau^{}
= 1\!:\! y^{\frac{1}{4}}\!:\!0$\,,\,
with {$\,y^{\frac{1}{4}} \simeq 1.2\!-\!2.4\,$}.\,
In general, including the possible decay contribution of $\tau$ leptons,
we infer the following condition for DM couplings:
\beqa
\label{eq:mutau}
\la_e^{} \!:\! \(\la_\mu^4 \!+\!\frac{1}{6}\la_\tau^4\)^{\!\!\frac{1}{4}}
\,=\, 1\!:y^{\frac{1}{4}}\,,\,
\eeqa
where the flavor ratio $\,y$\, is confined into a rather narrow range
$\,y^{\frac{1}{4}} \simeq 1.2\!-\!2.4\,$ (90\%\,C.L.).
We may consider that the lepton-DM portal sector poses a
$\,\mu\!-\!\tau$\, flavor symmetry
and thus realizes $\la_\mu^{}\!=\la_\tau^{}$.\,
Then, we can use this to derive a neat coupling relation
from Eq.\eqref{eq:mutau},
\beqa
\lambda_e^{}\!:\! \lambda_\mu^{} \!:\! \lambda_\tau^{}
= 1\!:\! \tilde{y}^{\frac{1}{4}}\!:\!\tilde{y}^{\frac{1}{4}},
\eeqa
with
$\tilde{y}^{\frac{1}{4}}
\!=\!\(\frac{6}{7}y\)^{\frac{1}{4}}\!\simeq\! 1.2\!-\!2.3$\,.
This constraint provides important guideline for the DM model buildings.
It is encouraging to further apply this analysis for testing DM interactions
with leptons.

\vspace*{1mm}
\subsection{\,Impacts of DAMPE Future Running}	
\label{sec:future}
\label{sec:2.6}
\vspace*{1mm}

\begin{figure*}[t]
\centering
\vspace*{-1mm}
\includegraphics[height=6.0cm]{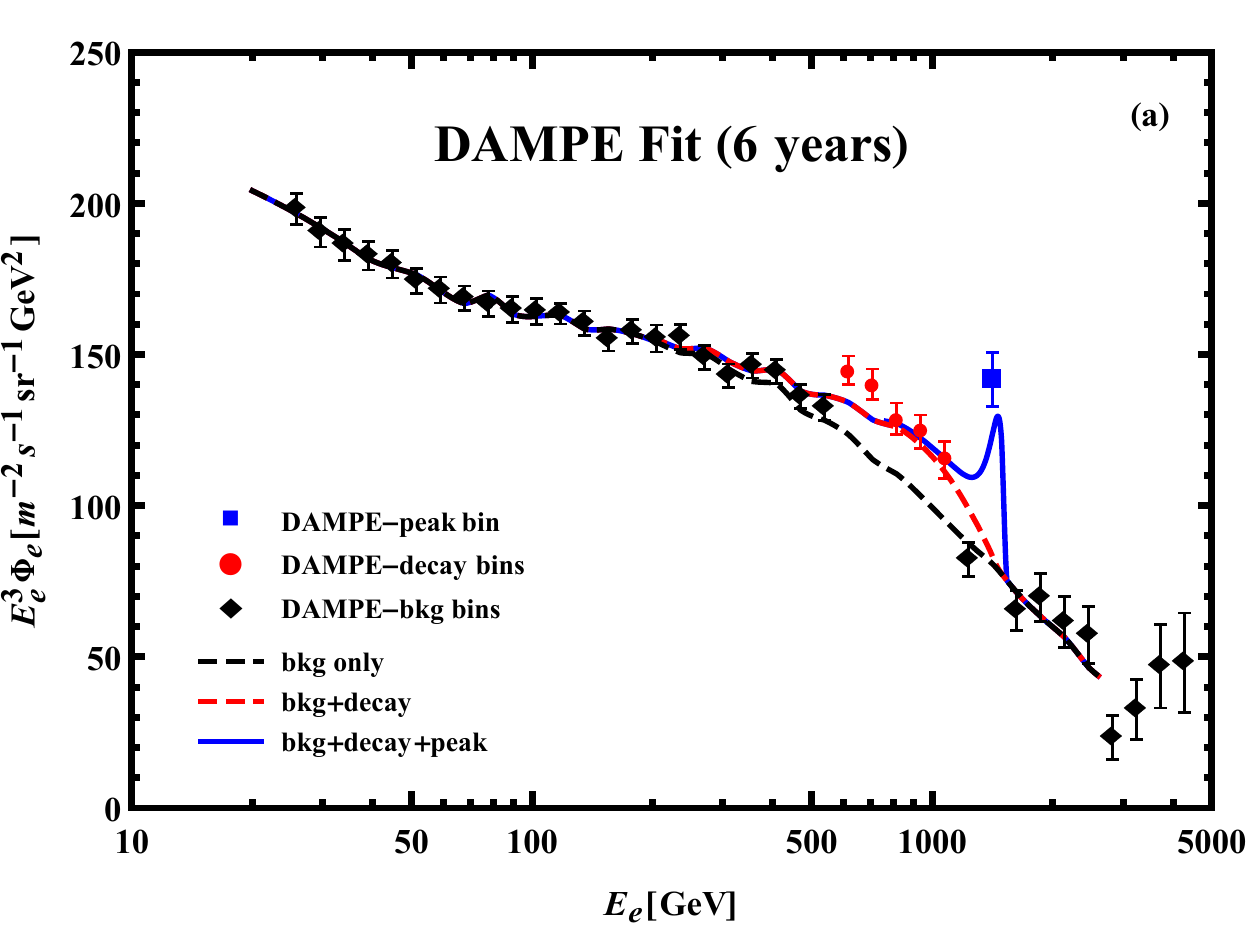}
\hspace*{2mm}
\includegraphics[height=5.8cm]{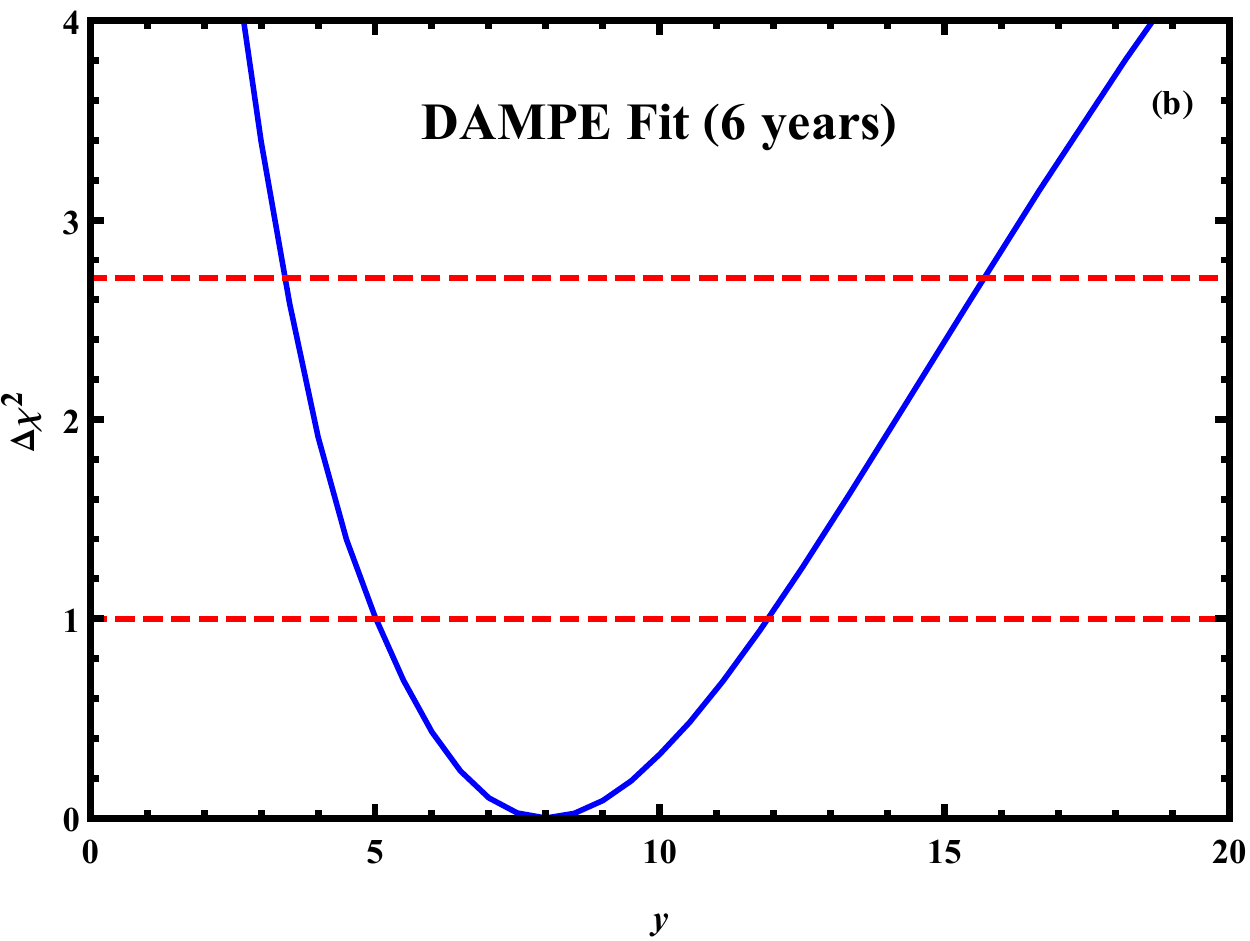}
\\[2mm]
\includegraphics[height=6.0cm]{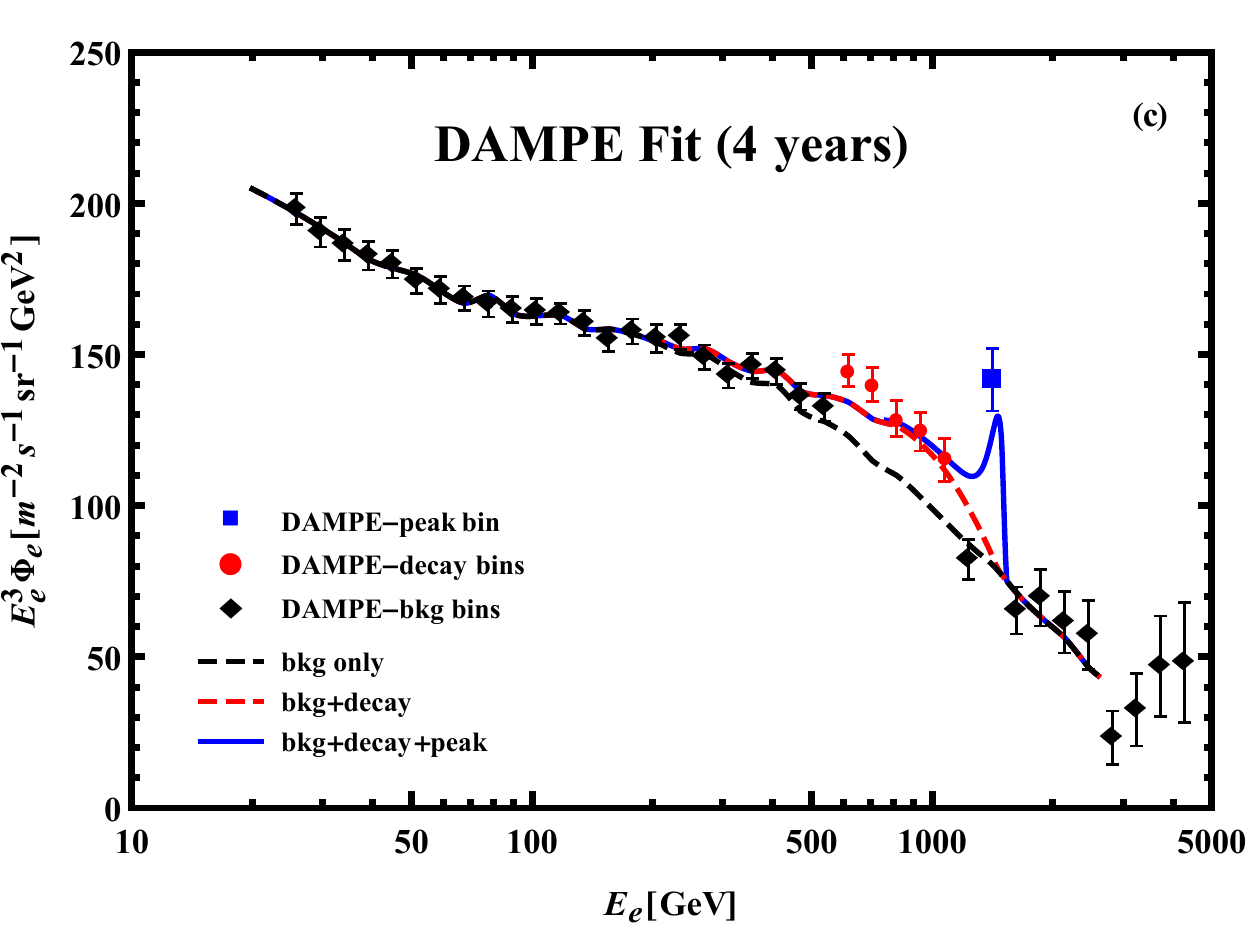}
\hspace*{2mm}
\includegraphics[height=5.8cm]{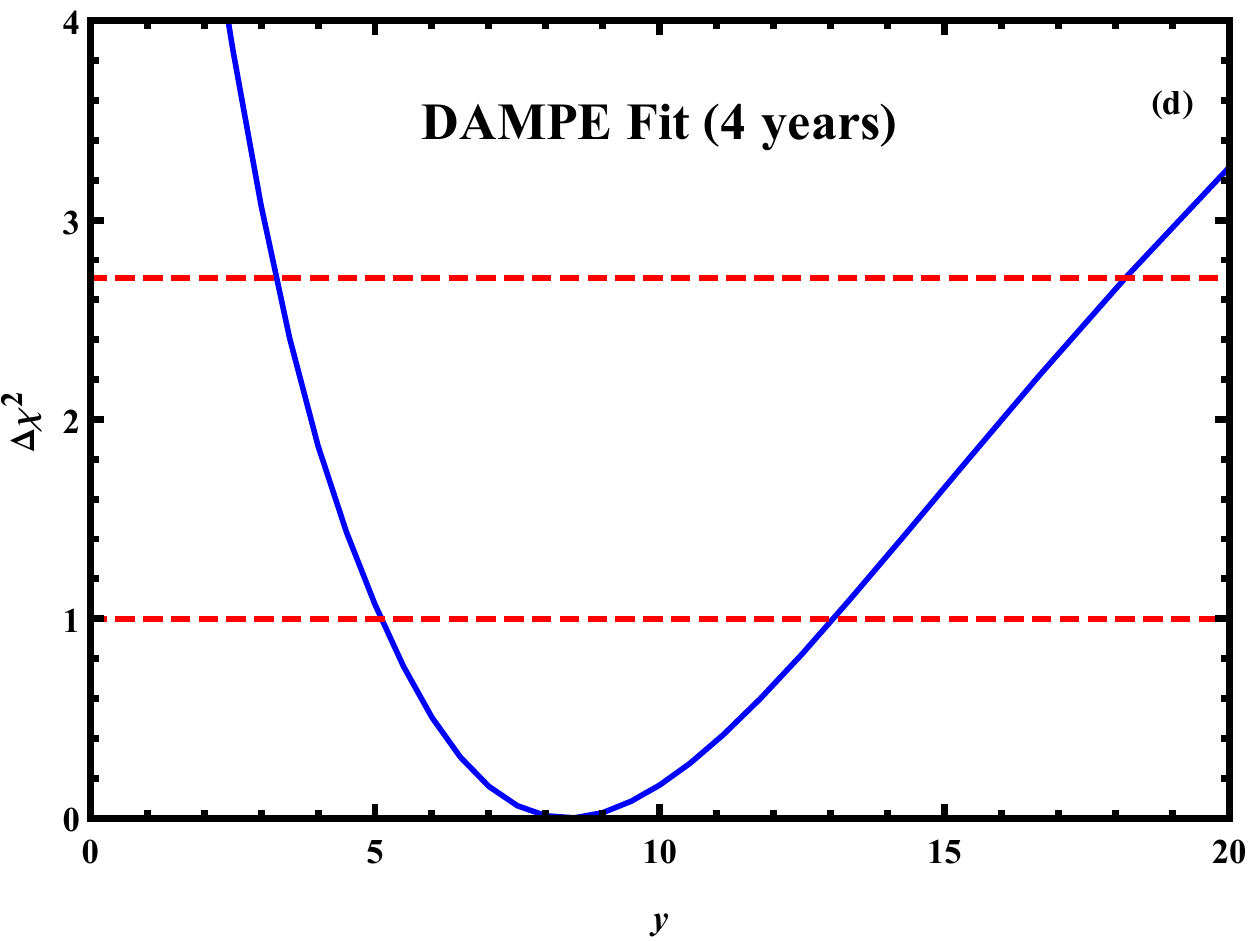}
\vspace*{-2mm}
\caption{\small
Fits of DAMPE CRE spectra for projected 4-year and 6-year data taking.
Plot\,(a): Fitting the DAMPE data with the projected 6-year data taking,
which reduces the current statistical errors by 50\%.
(Accordingly, the error bars are smaller than those in Fig.\,\ref{fig:3}(a)
by about a factor of $1/2$ at high energies.)
Plot\,(b): Marginalization of $y$ for the projected 6-year data taking of DAMPE,
which gives the constraints on the flavor ratio
$\,y=8.0^{+3.9}_{-3.0}$ ($1\sigma$) and $3.4<y<15.7$\, (90\%\,C.L.).
Plot\,(c): Fitting the DAMPE data with the projected 4-year data taking.
Plot-(d): Marginalization of $y$ for the projected 4-year data taking of DAMPE,
which gives the bounds $\,y=8.4^{+4.6}_{-3.3}$\, ($1\sigma$) and
$\,3.3<y<18.2$\, (90\% C.L.).
}
\label{fig:DAMPE-6y}
\label{fig:4}
\end{figure*}

Up to the end of 2019, the DAMPE satellite had accumulated data for 4 years,
and is expected to run up to $6$ years or even longer.
As the data size grows, the statistical errors will decrease.
Assuming the same central values of the detected CRE fluxes,
the statistical errors will scale with time as
$\,\sigma_\text{stat}^{}\!\propto t^{-1/2}$.\,
This means that the statistical errors of the 4-year data is reduced to
$\sqrt{1.5/4\,}\simeq 61\%$\, of that of the released data in 2017\,\cite{DAMPE2017},
and the 6-year measurements will reduce the statistical errors to
$\sqrt{1.5/6\,}=50\%$\,.\,
We assume the same systematic errors as before.

\vspace*{1mm}

We redo the fit for the projected 6-year data taking of DAMPE,
and present the results in Fig.\,\ref{fig:4}(a).
This new fit gives:
\beqa
A_e^{}\!=0.106\,\text{GeV}^{-1},~~~
\gamma_2^{}=2.73,~~~
\gamma_3^{}=2.27,~~~
E_\text{br2}^{}=57\,\text{GeV},~~~
E_\text{cut}^{}\!=2.46\,\text{TeV},
\hspace*{8mm}
\eeqa
for the background parameters,
together with the DM mass $\,M_\chi^{}\!\!=\!1.54\,$TeV,\,
and the DM annihilation cross sections
$\,\left< \sigma v \right>_e^{}
\!=\! 1.52\!\times\!10^{-26}\text{cm}^3\!/\text{s}$\,
and
$\,\left<\sigma v \right>_{\mu\tau}^{} \!=\! 1.22\!\times\!10^{-25}\text{cm}^3\!/\text{s}$\,.
For the fitting quality, we have
$\,\chi^2/\text{d.o.f.}\!=51.6/26=\!1.98$\,.
The increased $\chi^2$ value of this fit is mainly due to the reduction
of the statistical errors of the 6-year data taking.
The $\Delta\chi^2$ as a function of the flavor ratio $\,y\,$ is plotted
in Fig.\,\ref{fig:4}(b).
We obtain the flavor ratio
$\,y=8.0^{+3.9}_{-3.0}$ ($1\sigma$) and
$\,3.4\!<\!y\!<\!15.7$\, at 90\%\,C.L.
For the case of $\,y\!=\!0$\,,\, we find that $\chi^2$ increases by
$\,\Delta\chi^2=\chi^2(y\!=\!0)\!-\!\chi^2_{\min}\!=11.1$,\,
in which the 5 red bins provide $\,\Delta\chi^2\!=12.4$\,
as the dominant contribution.
The reason why the total increase $\Delta\chi^2$  is less than
the contribution from the 5 red bins
is because in certain bins the changes in $\Delta\chi^2$ are negative.

\vspace*{1mm}

For comparison, we apply the same analysis to the projected 4-year DAMPE
data. We present the results in Fig.\,\ref{fig:4}(c)-(d).
This fit gives the DM mass $\,M_\chi^{}\!=\!1.54\,$TeV,
and annihilation cross sections
$\,\left< \sigma v \right>_{\mu\tau}^{}\!\! =\! 1.28\!\times\!10^{-25}\text{cm}^3\!/\text{s}$\,
and
$\,\left< \sigma v \right>_e^{}\!\! =\! 1.53\!\times\!10^{-26}\text{cm}^3\!/\text{s}$\,.\,
From Fig.\,\ref{fig:4}(d), we derive bounds on the flavor ratio,
$\,y=8.4^{+4.6}_{-3.3}$\, at the $1\sigma$ level and
$\,3.3<y<18.2$\, at the 90\%\,C.L.
This fit also gives the corresponding background parameters,
\beqa
A_e^{}\!=0.116\,\text{GeV}^{-1},~~~
\gamma_2^{}=2.76,~~~
\gamma_3^{}=2.28,~~~
E_\text{br2}^{}=52\,\text{GeV},~~~
E_\text{cut}^{}\!=2.50\,\text{TeV}.
\hspace*{8mm}
\eeqa
Its fitting quality is given by 
$\chi^2/\text{d.o.f.}=39.6/26=1.52$\,.

\vspace*{1mm}
\section{Comparison with Other CRE Detections}
\label{sec:comp}
\label{sec:3}
\vspace*{1mm}

In this section, we extend our analysis of Section\,\ref{sec:2}
to further combine the CRE measurement of Fermi-LAT\,\cite{Fermi} in Section\,\ref{sec:3.1},
and then compare it with the recent data from
AMS-02\,\cite{AMS} and CALET\,\cite{CALET}
in Section\,\ref{sec:3.2}.

\subsection{\,Combined Fit with Fermi-LAT}	
\label{sec:Fermi-LAT}
\label{sec:3.1}
\vspace*{1mm}

We note that the CRE measurement by the Fermi-LAT collaboration\,\cite{Fermi} 
shows good consistency with
the DAMPE result\,\cite{DAMPE2017} even for the high energy region.
We observe that the CRE spectrum of Fermi-LAT exhibits a rise
around $(0.6\!-\!1.2)$\,TeV region,
which is consistent with the non-resonant excess (5 red bins) of the DAMPE result
in Fig.\,\ref{fig:1}, and thus may hint the $\mu^\pm$ ($\tau^\pm$) decay signals.
Although the Fermi-LAT data do not show a clear peak-like signal around $(1.3-1.5)$\,TeV
(unlike the DAMPE data), 
this is still consistent because the energy resolution of Fermi-LAT detector 
is about $(10\!-\!20)\%$ \cite{Fermi} above 1\,TeV energy scale,
and is much larger than the DAMPE energy resolution
$(1\!-\!2)\%$ \cite{DAMPE2017,DAMPE}.

\begin{figure*}[t]
\centering
\vspace*{-1mm}
\includegraphics[height=6cm]{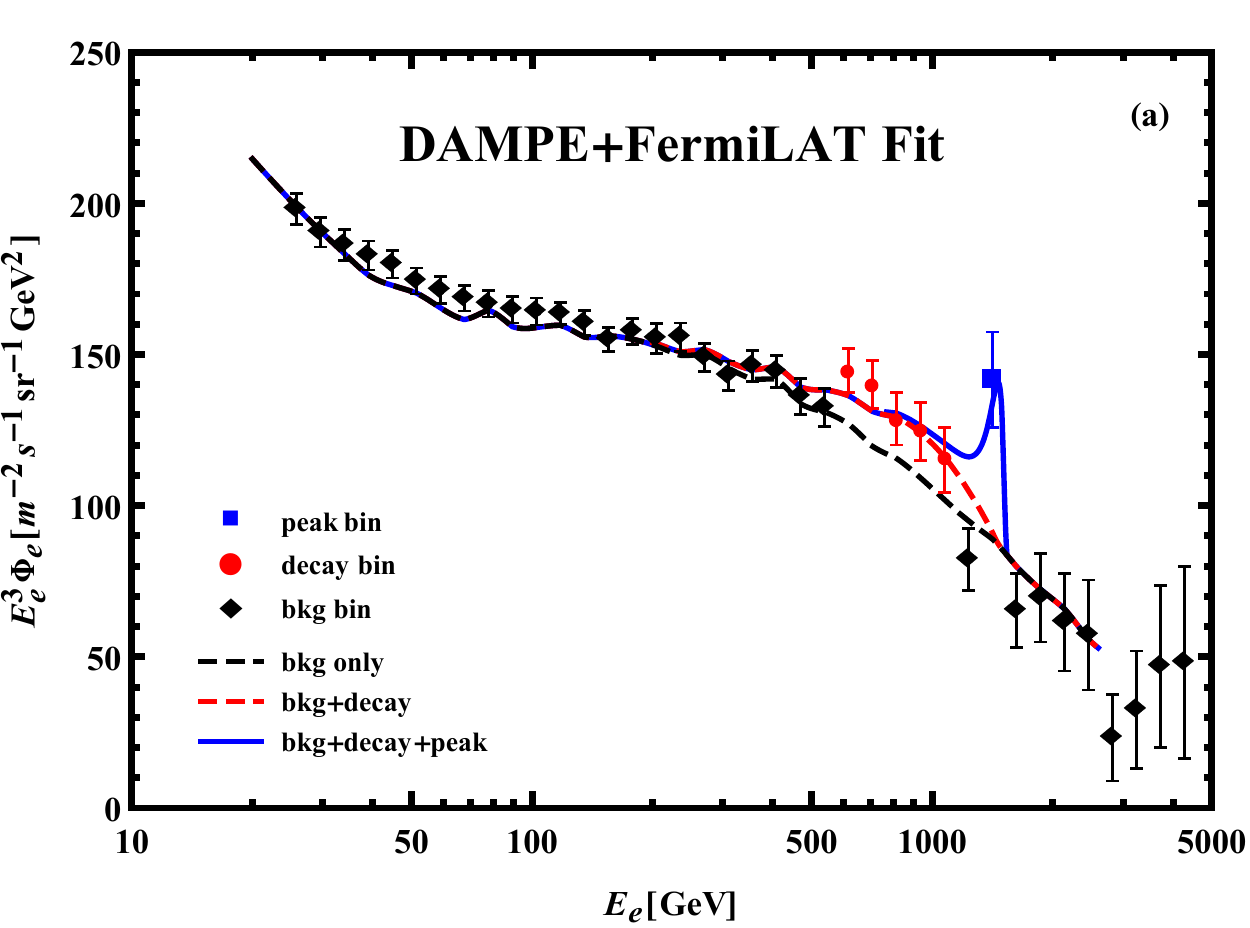}
\hspace*{3mm}
\includegraphics[height=6cm]{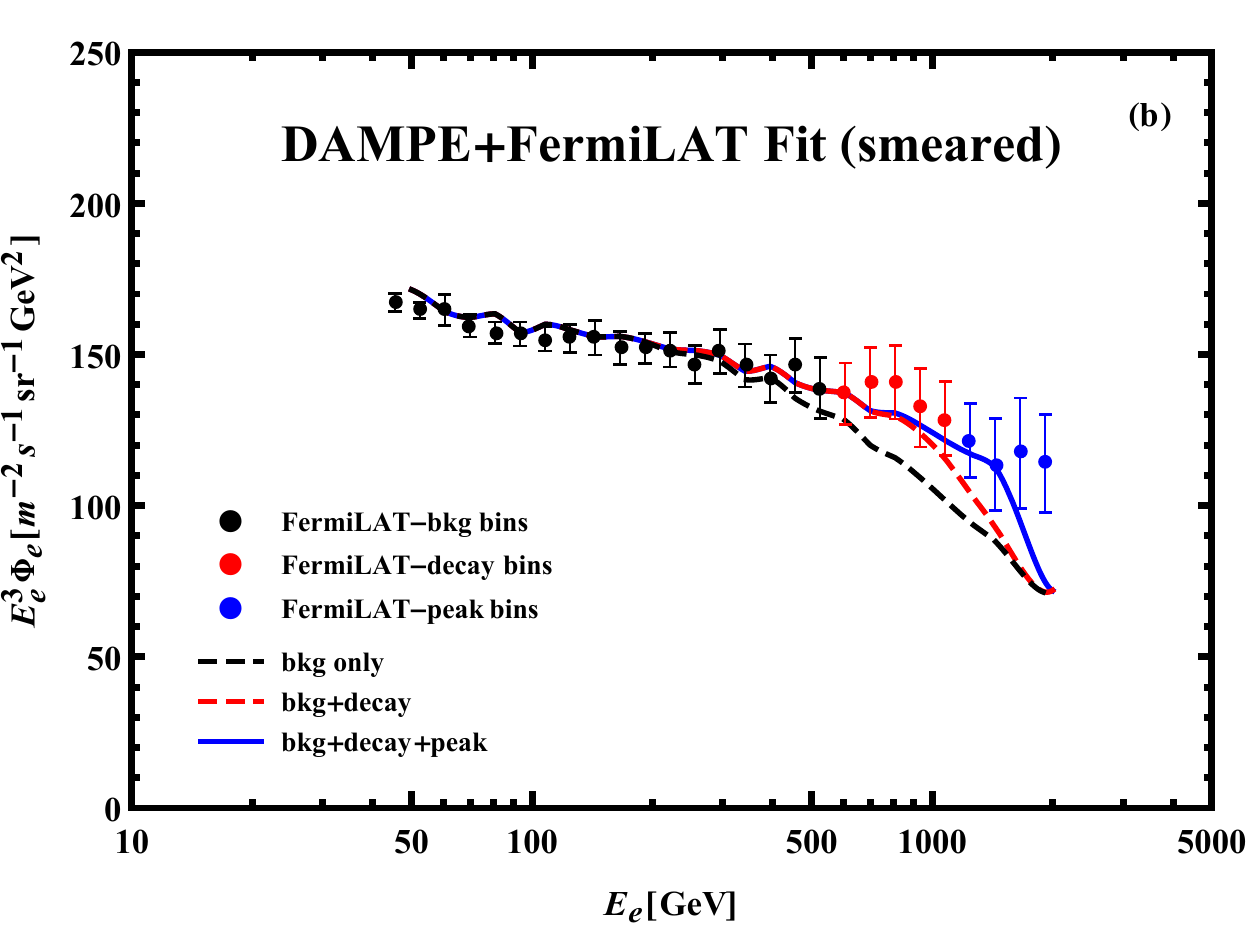}
\\[2mm]
\includegraphics[height=5.8cm]{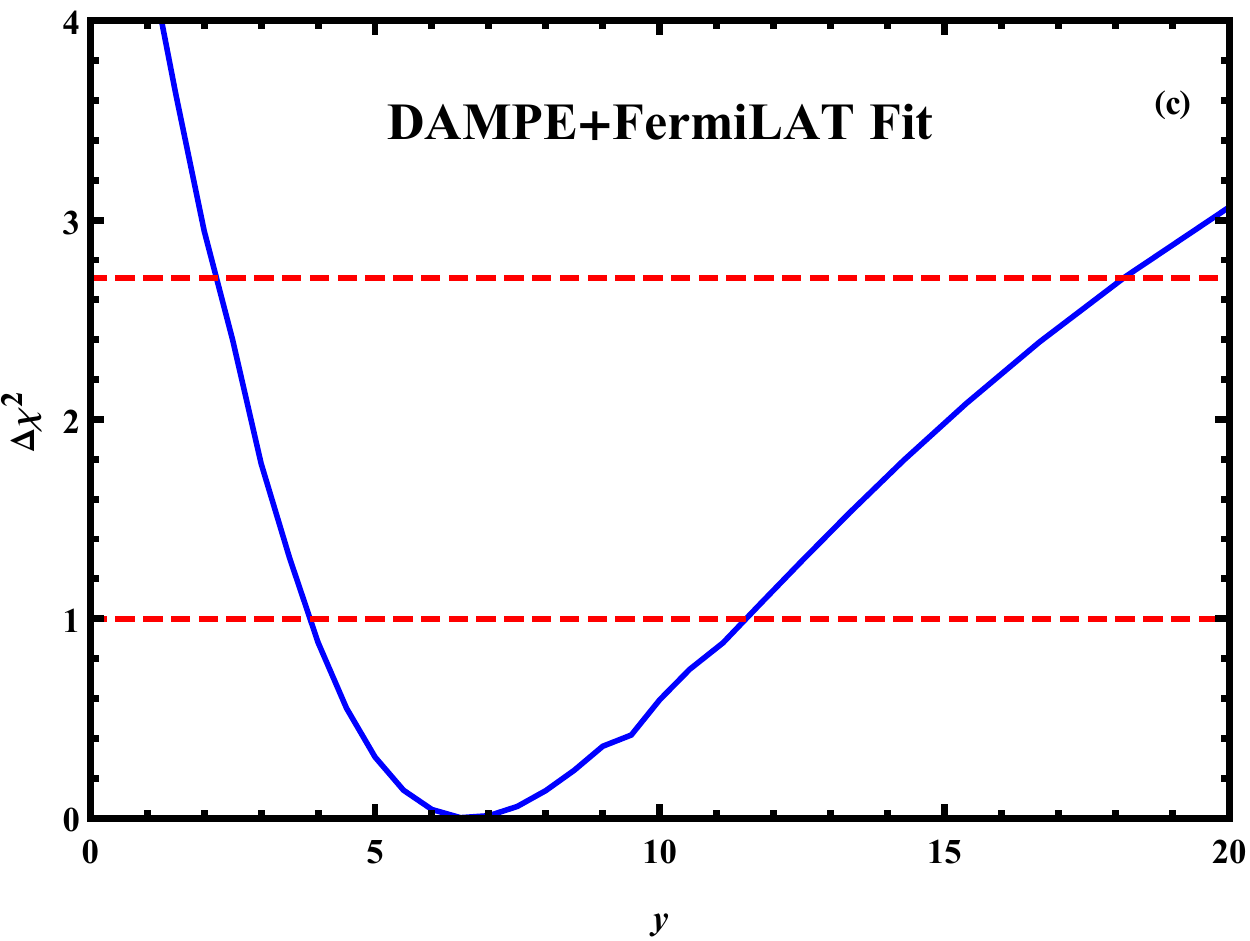}
\vspace*{-2mm}
\caption{\small
Combined fit of CRE energy spectra for both DAMPE and Fermi-LAT data.
Plot\,(a): The best-fit result is compared with the DAMPE data points
(with error bars).
Plot\,(b): The same best-fit result is convolved with Fermi-LAT's
energy resolution (marked as ``smeared''),
and is compared with the Fermi-LAT data points (with error bars).
Plot\,(c): The $\Delta\chi^2$ values versus $y$ from our fit,
which gives $\,y=6.6^{+4.9}_{-2.8}$\, at $1\sigma$ and $2.2<y<18.1$ at 90\% C.L.
}
\label{fig:DAMPE+Fermi-LAT}
\label{fig:5}
\end{figure*}

\vspace*{1mm}

In the following, we will make a combined fit for
both the DAMPE data (25\,GeV$-$2.6\,TeV)
and the Fermi-LAT data with the high energy (HE) selection (42\,GeV$-$2.1\,TeV).
For this, we define the total $\chi^2$ as follows,
\begin{eqnarray}
\label{eq:DFchi2}
\chi^2 & = & \chi^2(\text{DAMPE})+\chi^2(\text{Fermi-LAT})
\nn \\[1.5mm]
& = & \sum_j \!\left[\frac{1}{\,\sigma_j^\text{D}\,}
\!\(\!\Phi_j^\text{D}-\int_{E_j^{l\text{D}}}^{E_j^{u\text{D}}}\!\!\!
\frac{\Phi_e^\text{D}(E)}{\,E_j^{u\text{D}}-E_j^{l\text{D}}\,}\,\d E\)\right]^{\!2}
+ (\text{D}\leftrightarrow\text{F})\,,
\end{eqnarray}
where the superscript D (F) stands for DAMPE (Fermi-LAT),
and the quantities $E_j^{}$, $E_j^l$, $E_j^u$, $\sigma_j^{}$ and $\Phi_j^{}$
are the same as those defined in Eq.(\ref{eq:Dchi2}).

\vspace*{1mm}

The DAMPE detector has high energy resolution of
$(1\!-\!2)\%$ \cite{DAMPE2017,DAMPE},
so we have $\,\Phi_e^\text{D}\simeq \Phi_e^{}\,$ for DAMPE data,
where $\Phi_e^{}$ is the original CRE flux spectrum.
But for Fermi-LAT data, {because} the energy resolution is much lower than
DAMPE, we should take it into account for the measured flux $\Phi_e^\text{F}$.\,
The energy resolution can be described by a Gaussian distribution.
For the Fermi-LAT data,
we convolve $\Phi_e^{}$
with this Gaussian distribution through the integral,
\begin{eqnarray}
  \Phi_e^\text{F}(E^\text{F}) & \!\!\equiv\!\!\! &
  \int\! \Phi_e^{}(E)\frac{1}{\sqrt{2\pi}\Delta_E}
  \exp\!\left[-\frac{(E^\text{F}\!\!-\!E)^2}{2\Delta_E^2}\right]\! \d E\,,
\end{eqnarray}
where $\Delta_E^{}$ is the energy resolution at a given energy $E$\,.
The specific values of $\Delta_E^{}$ (as a function of $E$) are
adopted from Fig.\,10 of Ref.\,\cite{Fermi}.

\vspace*{1mm}

We fit the DAMPE data and Fermi-LAT data by minimizing the combined $\chi^2$.\,
The quality of the best fit is
$\,\chi^2/\text{d.o.f.}\!=61.0/53\simeq 1.15$\,.
The fit gives the background parameters,
\beqa
A_e^{}=0.762\,\text{GeV}^{-1},~~~
\gamma_2^{}=3.42,~~~ \gamma_3^{}=2.28,~~~
E_\text{br2}^{}=28\,\text{GeV},~~~
E_\text{cut}^{}=3.09\,\text{TeV},
\eeqa
as well as the DM mass $\,M_\chi^{}\!=\!1.54\,$TeV,
and annihilation cross sections
$\,\left< \sigma v \right>_{\mu\tau}^{}\!\!=\!
 1.07\!\times\!10^{-25}\text{cm}^3\!/\text{s}$\,
and
$\,\left< \sigma v \right>_e^{}
\!=\! 1.61\!\times\!10^{-26}\text{cm}^3\!/\text{s}$\,.
In Fig.\,\ref{fig:DAMPE+Fermi-LAT}, we present our fit of
both fluxes $\Phi^\text{D}$ (for DAMPE data) and
$\Phi^\text{F}$ (for Fermi-LAT data) in the plots\,(a) and (b), respectively.
These arise from the same fit, but due to the convolution with the energy resolution
of Fermi-LAT, the shape of the curve of $\Phi^\text{F}$ [plot\,(b)] is not identical
to that of $\Phi^\text{D}$ [plot\,(a)].

\begin{table*}[t]
	\centering
	\begin{tabular}{c||c|c|c|c|c}
		\hline\hline
		&&&&& \\[-3.5mm]
		Experiments & $\chi^2\!/\text{d.o.f}$ & $M_\chi^{}$\,(TeV)
		& $\langle\sigma v \rangle_{\!e}^{}$
		& $\langle\sigma v\rangle_{\!\mu\tau}^{}$ & $y$~
		\\[1mm]
		\hline\hline
		&&&&& \\[-3.8mm]
		DAMPE & 19.6/26=0.756 & 1.54 & 0.154 & 1.39 & $9.0_{-4.2}^{+8.4}$~
		\\[0.7mm]
		\hline
		&&&&& \\[-3.8mm]
		DAMPE (in 4y) & 39.6/26=1.52 & 1.54 & 0.153 & 1.28 & $8.4_{-3.3}^{+4.6}$~
		\\[0.7mm]
		\hline
		&&&&& \\[-3.8mm]
		DAMPE (in 6y) & 51.6/26=1.98 & 1.54 & 0.152 & 1.22 & $8.0_{-3.0}^{+3.9}$~
		\\[0.7mm]
		\hline
		&&&&& \\[-3.8mm]
		DAMPE+Fermi-LAT & 61.0/53=1.15 & 1.54 & 0.161 & 1.07 & $6.6_{-2.8}^{+4.9}$~
		\\[0.7mm]
		\hline\hline
	\end{tabular}
\caption{\small
Comparison of the fitting results for DAMPE and Fermi-LAT experiments.
The second row presents our fit of the current DAMPE data, the third (fourth)
row shows the fit for the projected 4-year (6-year) data taking of DAMPE,
and the fifth row gives our combined fit of the current DAMPE and Fermi-LAT data.
The fourth and fifth columns summarize the best fits of the DM annihilation
cross sections via $e^\pm$ and $\mu^\pm$\,($\tau^\pm$) channels, respectively
(in unit of $10^{-25}\text{cm}^3\!$/s). The last column shows
the central value and the $1\sigma$ bounds of the flavor ratio
$\,y=\langle\sigma v\rangle_{\!\mu\tau}^{}/\langle\sigma v \rangle_{\!e}^{}\,$.
}
\label{tab:2}
\vspace*{2mm}
\end{table*}

\vspace*{1mm}

Furthermore, we plot $\Delta\chi^2$ as a function the flavor ratio $\,y\,$
in Fig.\,\ref{fig:DAMPE+Fermi-LAT}(c). From this, we derive the
$1\sigma$ bound on the flavor ratio $\,y=6.6^{+4.9}_{-2.8}$,
and the 90\% C.L. bound, $2.2<y<18.1$\,.
We note that the case of $\,y=0$\, worsens the fit quality considerably,
with the $\chi^2$ value increasing by $5.9$\,.
This indicates again that including the $\mu^\pm$\,($\tau^\pm$)
decay contribution is favored
by the combined data of DAMPE and Fermi-LAT.

\vspace*{1mm}

Finally, for comparison, we summarize in Table\,\ref{tab:2}
our fits of the CRE energy spectra for
the current DAMPE data, the projected 4-year and 6-year running of DAMPE,
and the combined current data set of DAMPE and Fermi-LAT experiments.

\vspace*{1mm}
\subsection{\,Comparison with AMS-02 and CALET}	
\label{sec:CALET}
\label{sec:3.2}
\vspace*{1mm}

In this subsection, we will further compare our DAMPE analysis
(Section\,\ref{sec:2}) with the AMS-02 and CALET data.

\vspace*{1mm}

The CALET collaboration gave a recent update of its CRE measurement
over the energy range $(0.011\!-\!4.8)$TeV in June,\,2018\,\cite{CALET}.
The CRE spectrum of CALET is consistent with a break at 0.9\,TeV (as found
by the DAMPE observation\,\cite{DAMPE2017}). The CALET data do not show a peak structure
around $(1.3\!-\!1.5)$\,TeV,  but it is premature to draw any firm conclusion
since CALET still has sizable statistical uncertainties beyond 1\,TeV energy.
%
%
The AMS collaboration published new precision measurements of
cosmic ray electrons up to 1.4\,TeV in March, 2019\,\cite{AMS}.
Together with their positron measurement published in January,\,2019,
the CRE spectrum of AMS-02 shows agreement with the CALET result\,\cite{CALET}
up to 1\,TeV.
However, this measurement does not reach the energy region of the peak excess
shown by DAMPE data.
We can hardly compare the non-peak excess of DAMPE with the AMS-02 result as well,
because the latter only shows two data points over the energy region
$(0.6-1.1)$\,TeV.

\vspace*{1mm}

In view of the current situation discussed above,
we need not to fit all the DM parameters with the AMS-02 and CALET data
in this subsection.
Instead, we will just test a few benchmark cases motivated by our analysis
in Section\,\ref{sec:2.4}.
We combine the CALET data (using the same energy binning as DAMPE analysis)
with the AMS-02 data (electron\,+\,positron),
and apply the $\chi^2$ fit to the following three cases:

\begin{figure}[t]
	\centering
	\vspace*{-1mm}
	\includegraphics[height=6cm]{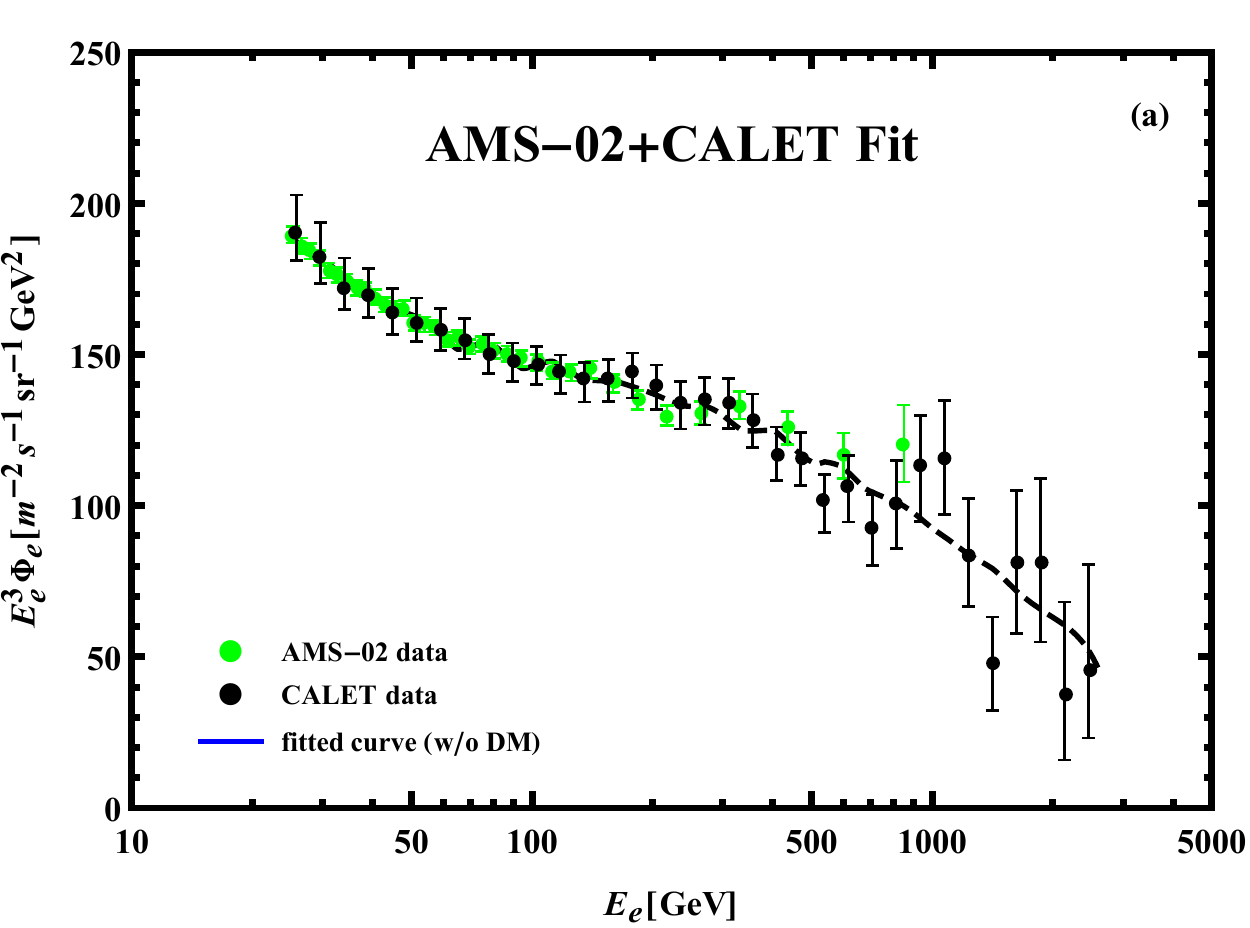}
	\hspace*{3mm}
	\includegraphics[height=6cm]{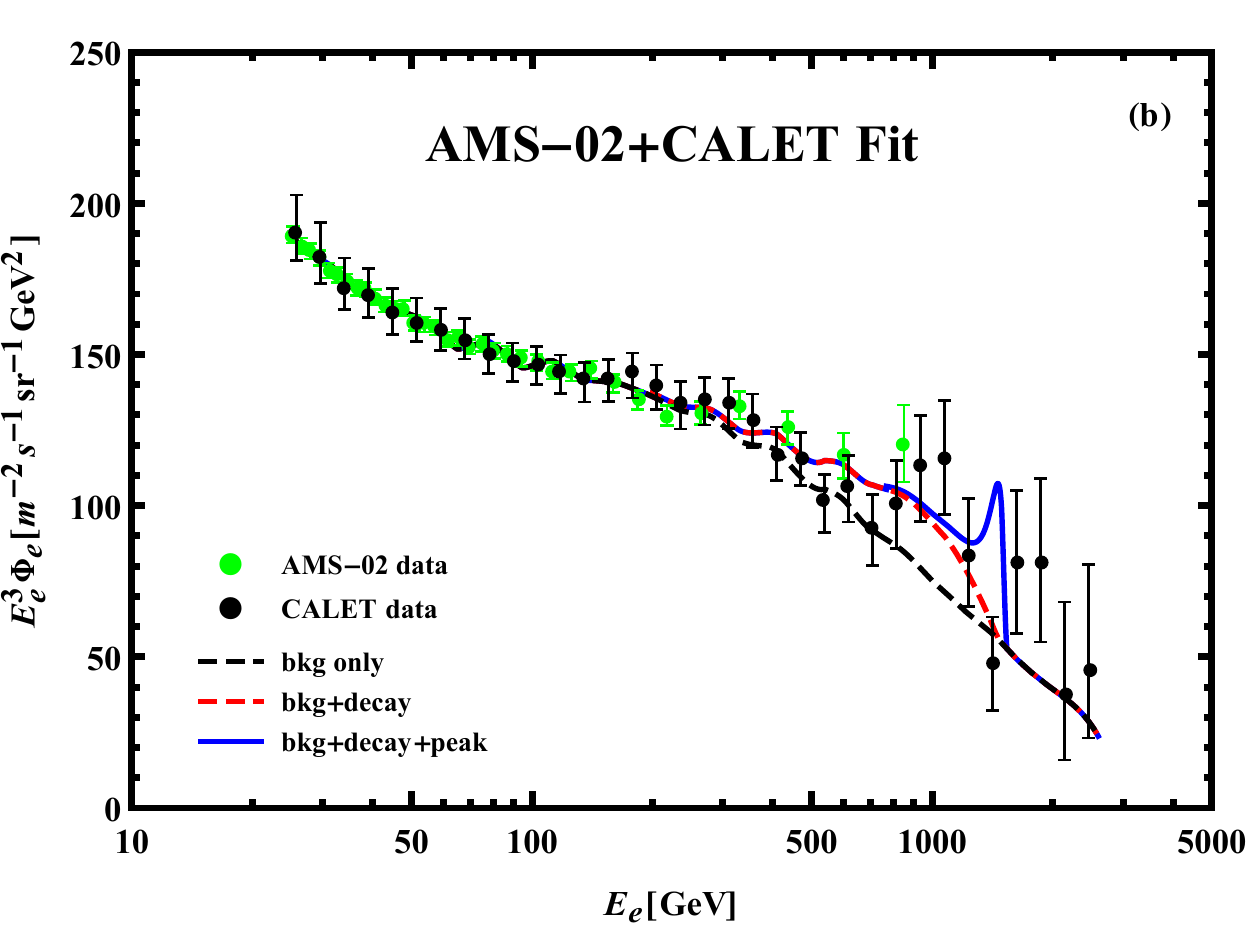}
	\\[2mm]
	\includegraphics[height=6cm]{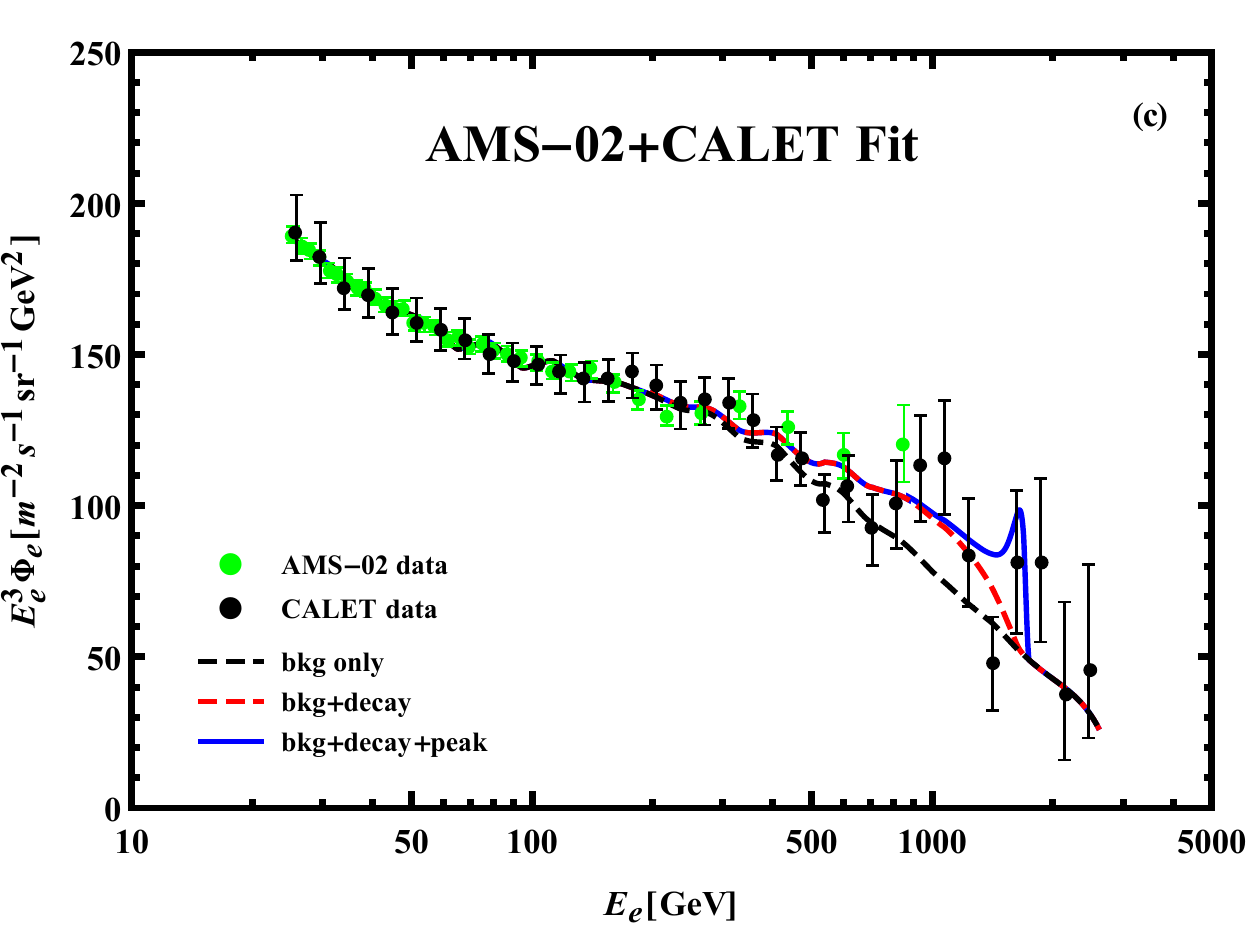}
	\vspace*{-2mm}
	\caption{\small
		Combined fit of the CRE energy spectra from AMS-02 and CALET.
		In each plot, the AMS-02 (CALET) data are shown as green (black) dots,
		including their $1\sigma$ error bars.
		Plot\,(a): The best fit without DM contribution, defined as Case-(a) in the text,
		with $\,\chi^2\!/\text{d.o.f.}\!=\!36.32/64=0.567$.\,
		Plot\,(b): The best fit of CALET in Case-(b)
		with $\chi^2\!/\text{d.o.f.}=43.95/61=0.721$,\,
		where we input the best-fit of the flavor ratio $y$ from DAMPE data
		as in Section\,\ref{sec:2.4}.
		Plot\,(c): The best fit of Case-(c) which includes the DM mass for fit and gives
		$M_\chi^{}\!=1.745$\,TeV. The corresponding fit quality,
		$\chi^2/\text{d.o.f.}\! =37.39/61=0.613$,
		is comparable to Case-(a).
	}
	\label{fig:CALET}
	\label{fig:6}
	\vspace*{2mm}
\end{figure}
\begin{enumerate}
\vspace*{-3mm}
\item[(a).]
There is only background CRE flux
without assuming any $e^\pm$ signals from the DM annihilation.

\vspace*{-3mm}
\item[(b).]
There exist $e^\pm$ signals from the DM annihilations,
with the DM mass $M_\chi^{}\!=1.535$\,TeV, the annihilation cross section
$\left<\sigma v \right>_e^{}\!=1.54\!\times\!10^{-26}\text{cm}^3\!/\text{s}$\,
and the flavor ratio $\,y=9.0\,$ as given by the best fit of Section\,\ref{sec:fit}.

\vspace*{-3mm}
\item[(c).]
Same as Case-(b), but the DM mass is allowed to vary within the range
$\,M_\chi^{} \!\!\in\! [1.7, 1.9]$\,TeV.

\vspace*{-3mm}
\end{enumerate}
For the above three cases, our fits have the fit quality
\beqa
\chi^2\!/\text{d.o.f.} = (36.32/64,\,43.95/61,\,37.39/61) \simeq
(0.567,\, 0.721,\, 0.613),\,
\eeqa
respectively.
Also, the fit of Case-(c) gives the DM mass $\,M_\chi^{}\!=1.745$\,TeV.
We see that adding the dark matter contribution does not affect much the fitting quality.
We present the fitted CRE energy spectra for the three cases
in plots (a)-(c) of Fig.\,\ref{fig:CALET}.

\vspace*{1mm}

Inspecting the CALET CRE spectrum alone,
we note that there apppear two broad excesses around the windows
$(0.75\!-\!1.2)$\,TeV and $(1.5\!-\!2.0)$\,TeV.
The first one is not only compatible with AMS-02 data,
but also coincide with the non-peak excess range of DAMPE as we uncovered
in Fig.\,\ref{fig:1} (the 5 red bins).
The second excess is around the energy window somewhat higher than
the peak excess of DAMPE (the blue bin in Fig.\,\ref{fig:1}),
but our fit shows that it can match a peak structure from the DM signals
with mass $M_\chi^{}\!=1.745$\,TeV.

\vspace*{1mm}

The current CRE spectra of DAMPE and CALET data show some different features,
especially at energies around 1.5\,TeV, which are yet to be resolved after collecting
much larger data samples in the near future. But such differences are not significant
given the much larger error bars of CALET data at energies above 1\,TeV.
Regardless of the unknown systematic effects,  our above analysis still shows
some compatible spectral features between these experimental data,
which may be the hint of certain dark matter annihilation processes nearby
and are worth of attention.
Hence, the upcoming data of the on-going DAMPE and CALET measurements
will be extremely helpful.

\vspace*{1.5mm}
\section{Gamma Ray Constraints from Fermi-LAT}
\label{sec:gamma}
\label{sec:4}
\vspace*{1mm}

In this section, we analyze constraints on the DM annihilation signals
from the $\gamma$-ray measurements of Fermi-LAT.
There have been some discussions on the possible constraints
from the $\gamma$-ray observations\,\cite{gamma}. Most of the literatures
assumed point-like sources of the $\gamma$-ray emission from the nearby
subhalo. But such a nearby subhalo is in fact a largely extended object,
so more dedicated search and analysis are necessary.
The $\gamma$-rays produced by the leptons from DM annihilations
consist of two main components: the prompt $\gamma$-rays and the secondary
$\gamma$-rays. The prompt $\gamma$-rays are emitted during the DM annihilations,
including the final state radiation and photons from the hadronic decays of
$\,\tau$\, leptons. The case of secondary $\gamma$-rays belongs to
the secondary emissions, mainly due to the Inverse Compton
Scattering (ICS) emissions of electrons/positrons when propagating in the
interstellar radiation field. For the nearby subhalo scenario considered
in the present study, the ICS $\gamma$-rays are generally
small and negligible as compared with the prompt $\gamma$-rays\,\cite{gamma},
although there is some dependence on the location of the subhalo.
Thus, we will focus on the prompt $\gamma$-rays in this analysis.

\vspace*{1.5mm}
\subsection{\,Prompt Gamma Rays}	
\label{sec:prompt}
\label{sec:4.1}
\vspace*{2mm}

\begin{figure}[t]
\centering
\includegraphics[height=6.0cm]{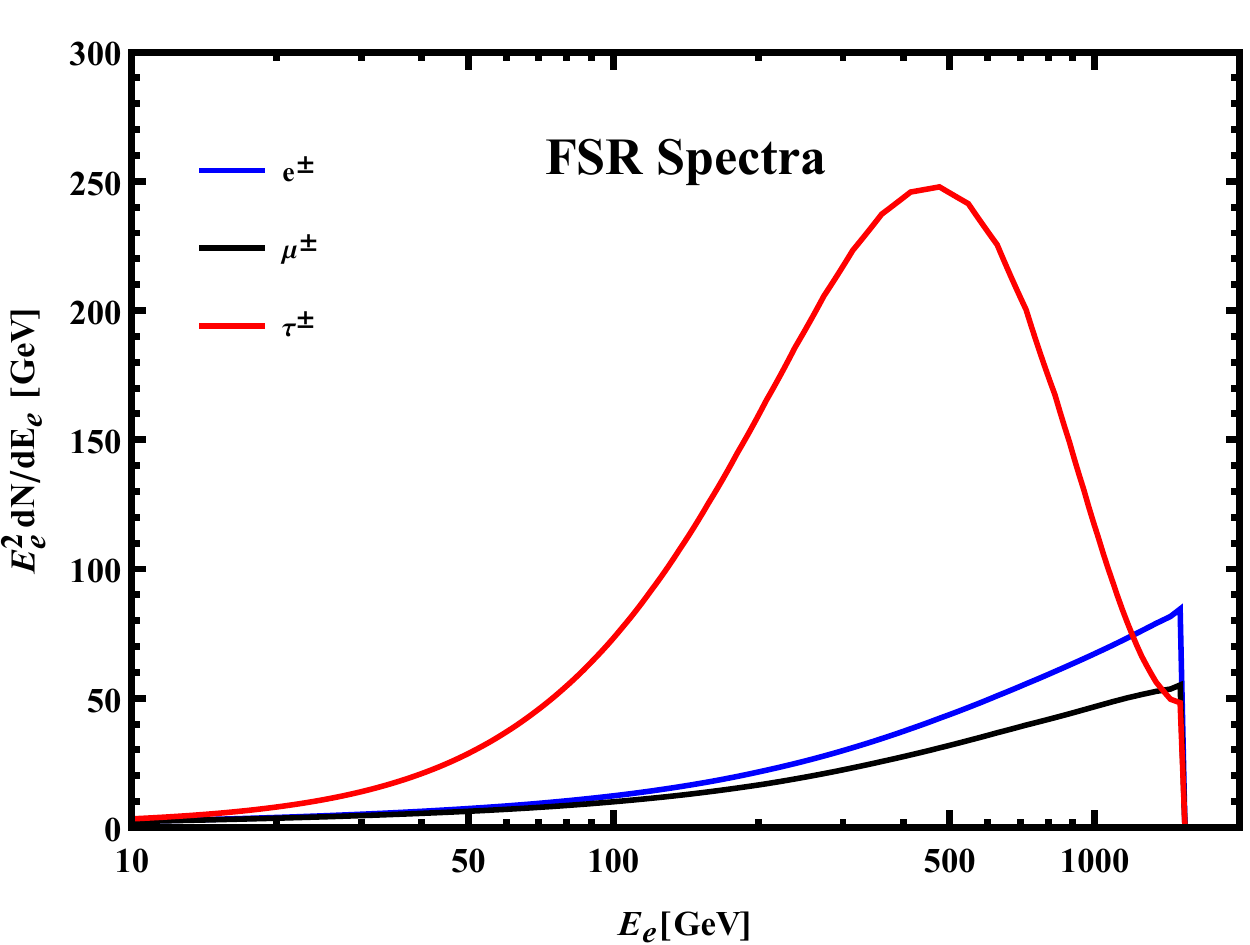}
\vspace*{-2mm}
\caption{\small
Final state radiation (FSR) spectra for $e^\pm$, $\mu^\pm$ and $\tau^\pm$
(with 1.54\,TeV energy), shown as (blue, black, red) curves, respectively.
The $\tau^\pm$ channel (red curve) dominates the sub-TeV
region, so its fraction will be strongly constrained.
}
\label{fig:FSR}
\label{fig:7}
\vspace*{3mm}
\end{figure}

We apply the PPPC4DMID code\,\cite{PPPC}\cite{Flux} to compute the energy
spectra of the prompt photon emissions,\footnote{%
There is another package for this calculation\,\cite{QCD}.
We find that the FSR photon spectra of \cite{QCD}
are in good agreement with PPPC4DMID, so we use the PPPC4DMID results
in the current study for consistency.} as shown in Fig.\,\ref{fig:7}.
The $\tau^\pm$ channel significantly differs from that of the $e^\pm$
and $\mu^\pm$ channels, because of the $\pi^0$ products from
$\tau^\pm$ decays. For the $e^\pm$ and $\mu^\pm$ channels, the
final state radiation (FSR) comes from
the internal bremsstrahlung radiation only.
In Sections\,\ref{sec:2} and \ref{sec:3}, we have determined the value
of the flavor ratio $\,y\,$,\,
which is a combination of the $\mu^\pm$ fraction and the $\tau^\pm$
fraction, through fitting to the CRE data. The $\gamma$-ray data are
expected to give further constraints on the $\tau^\pm$ fraction due
to its abundant photon production.

%

\vspace*{1mm}
\subsection{\,The $\mathbf{J}$ factor}	
\label{sec:jfactor}
\label{sec:4.2}
\vspace*{2mm}

The expected photon flux from the DM annihilations in a nearby subhalo
is given by
\begin{eqnarray}
\d\Phi_\gamma \,=\,
\frac{J}{\,16\pi M_\chi^2\,}
\left(\sum_{f}\langle\sigma v\rangle_f^{}\frac{\,\d N_f^{}\,}{\d E}\right)\d\Omega\d E \,.
\end{eqnarray}
In the above, the $J$ factor is defined as
\begin{eqnarray}
\label{eq：J-def}
J(\theta) \,=\, \int_\text{l.o.s.}^{}
\hspace*{-3mm}\d s\, \rho_\chi^2(r) \,,
\end{eqnarray}
where $\,\theta\,$ is the angle between the line-of-sight (l.o.s.) and the
subhalo center, $r$ is the radial distance from a point on l.o.s.\ to
the subhalo center, and
$\,s\!=\!\!\sqrt{r^2\!+d_s^2\!-2rd_s^{}\!\cos\theta\,}$
is the l.o.s.\ distance to the observer.

\vspace*{1mm}

For convenience, we fit the distribution of \,$\ln\!J$\,
as a function of $\ln\theta$,\, and obtain
\begin{eqnarray}
\label{eq:Jfit}
\ln [J(\theta)] \,=\,
C - 0.479507 \ln \theta - 0.12386 (\ln \theta)^2 - 0.0265383 (\ln \theta)^3,
\end{eqnarray}
where $J\,$ is in unit of $10^{20}$\,GeV$^2$cm$^{-5}$
and $\,\theta\,$ is in unit of degrees.
The constant $C$ is irrelevant here,
since we are going to use a normalized skymap in the data analysis.
The error of this parametrization as compared to the numerical integral
of Eq.\eqref{eq：J-def} is less than $2.5\%$ for the angular range
$\,\theta<20^{\circ}$.

\vspace*{1mm}

\begin{figure}[t]
\centering
\includegraphics[height=7cm]{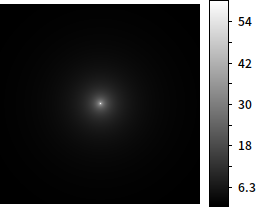}
\vspace*{1mm}
\caption{\small
Skymap of the normalized $J$ factor for the DM subhalo used in our CRE fits.
The vertical legend shows normalized J values with the corresponding colors
in the plot.
}
\label{fig:subhalo}
\label{fig:8}
\vspace*{3mm}
\end{figure}

We present in Fig.\,\ref{fig:8} the skymap of the normalized $J$ factor distribution
of the subhalo in a $40^\circ\!\times\!40^\circ$ region, which will be used
in the analysis of Fermi-LAT data in the next subsection.
In Table\,\ref{tab:3}, we list some typical values of the integrated
$J$ factor $\int \!\!J\d\Omega\,$ in the second row and
the averaged $J$ factor $\int\!\! J \d\Omega\,/\!\int\!\!\d\Omega$\,
in the third row.

\begin{table*}[h]
\vspace*{4mm}
	\centering
	\begin{tabular}{c||c|c|c}
		\hline\hline
		&&& \\[-3.5mm]
		Angular Radius $\theta_0^{}$
		& $0.1^\circ$ & $1.0^\circ$ & $10^\circ$
		\\[1mm]
		\hline\hline
		&&& \\[-3.5mm]
		Integrated $J$ Factor ($10^{20}$\,GeV$^2$cm$^{-5}$) & 2.819 & 146.5 & 3138
		\\[1mm]
		\hline
		&&& \\[-3.5mm]
		Averaged $J$ Factor ($10^{25}$\,GeV$^2$cm$^{-5}$sr$^{-1}$)
		& 2.946 & 1.531 & 0.3288
		\\[1mm]
		\hline\hline
\end{tabular}
\vspace*{2mm}
\caption{Typical values of the integrated and averaged $J$ factors.
	}
	\label{tab:3}
	\vspace*{1.5mm}
\end{table*}
%

\subsection{\,Gamma Ray Searches with Fermi-LAT}
\label{sec:extendedsource}
\label{sec:4.3}
\vspace*{2mm}

We use the Pass\,8 data recorded by Fermi-LAT
between August\,4, 2008 and July\,2, 2018
to search for possible $\gamma$-ray emission from such a DM subhalo.
We restrict our study with the CLEAN event class (evclass\,$=\!256$ and evtype\,$=\!3$).
To reduce the impact from the Earth limb, the events with zenith angles
$>\!90^\circ$ are excluded.
We perform the searches (in a series of sky directions)
within regions of interest (ROIs) centered at galactic coordinates
$(\ell,\,b)=(0^\circ,\,\pm 90^\circ)$, $(0^\circ,\,\pm 70^\circ)$,
$(90^\circ,\,\pm 70^\circ)$, $(180^\circ,\, \pm 70^\circ)$,
$(270^\circ,\, \pm 70^\circ)$, $(0^\circ,\, 0^\circ)$, $(0^\circ,\, 30^\circ)$,
and $(120^\circ,\, 45^\circ)$, respectively. The ROI radius is chosen to be
$20^\circ$.\, For most of ROIs, the energy range is chosen as
100\,MeV$-$500\,GeV. Since the expected signal appears mainly in the high
energy band, we perform the analysis in the range $(1\!-\!500)$\,GeV for ROIs
$(0^\circ,\,30^\circ)$ and $(120^\circ,\,45^\circ)$, and the range
$(10\!-\!500)$\,GeV for the ROI $(0^\circ,\, 0^\circ)$, in order to speed
up the analysis. For the $(0^\circ,\, 0^\circ)$ case, a higher energy
threshold can also reduce the impact from the reported Galactic center
GeV $\gamma$-ray excesses\,\cite{GCexcess}.

\vspace*{1mm}

We use the standard binned likelihood method with the science tool version
v10r0p5.\footnote{http://fermi.gsfc.nasa.gov/ssc/data/analysis/software/}\,
The model includes point sources from the 3FGL catalog\,\cite{Acero:2015hja},
the diffuse background templates gll\_iem\_v06 and
iso\_P8R2\_CLEAN\_V6\_v06,\footnote{%
http://fermi.gsfc.nasa.gov/ssc/data/access/lat/BackgroundModels.html}\,
as well as our postulated DM subhalo.
If no clear signal is found, the 95\% confidence upper limits (UL)
on the annihilation cross section are derived
(cf.\ Appendix\,\ref{app:A} for detail).
The mass of the DM particle is fixed to be $M_\chi^{}\!\!=\!1.54$\,TeV.
We study the following benchmark cases for the DM annihilation:
\begin{enumerate}
\vspace*{-1.5mm}
\item[(a).]
$\langle\sigma v\rangle_e=\langle\sigma v\rangle_\mu=\langle\sigma v\rangle_\tau=0$,
which corresponds to background only;

\vspace*{-3mm}
\item[(b).]
$\langle\sigma v\rangle_e=$ free,
$\langle\sigma v\rangle_\mu=\langle\sigma v\rangle_\tau=0$\,;

\vspace*{-3mm}
\item[(c).]
$\langle\sigma v\rangle_\mu=$ free,
$\langle\sigma v\rangle_e=\langle\sigma v\rangle_\tau=0$\,;

\vspace*{-3mm}
\item[(d).]
$\langle\sigma v\rangle_\tau=$ free,
$\langle\sigma v\rangle_e=\langle\sigma v\rangle_\mu=0$\,;

\vspace*{-3mm}
\item[(e).]
$\langle\sigma v\rangle_e=\langle\sigma v\rangle_\mu=\langle\sigma v\rangle_\tau=$ free,
which corresponds to $N_e:N_\mu:N_\tau=1:1:1$\,;

\vspace*{-3mm}
\item[(f).]
$\langle\sigma v\rangle_e=1.54\!\times\! 10^{-26}\text{cm}^3/\text{s}$,
$\langle\sigma v\rangle_\mu=$ free, $\langle\sigma v\rangle_\tau=0$\,;

\vspace*{-3mm}
\item[(g).]
$\langle\sigma v\rangle_e^{}=1.54\!\times\! 10^{-26}\text{cm}^3/\text{s}$,
$\langle\sigma v\rangle_\mu^{}=0$,
$\langle\sigma v\rangle_\tau^{}=$\,free\,;

\vspace*{-3mm}
\item[(h).]
$\langle\sigma v\rangle_e^{}=1.54\!\times\! 10^{-26}\text{cm}^3/\text{s}$,
$\langle\sigma v\rangle_\mu^{}=$ free,
$\langle\sigma v\rangle_\tau^{}=$\,free\,;

\vspace*{-3mm}
\item[(i).]
$\langle\sigma v\rangle_e=1.54\!\times\! 10^{-26}\text{cm}^3/\text{s}$,
$\langle\sigma v\rangle_\tau^{}=$ free,
$\langle\sigma v\rangle_\mu^{}=1.39\!\times\! 10^{-25}\text{cm}^3/\text{s}
 -0.178\langle\sigma v\rangle_\tau^{}$, which corresponds to
$\,y\equiv 9.0$\,.
\vspace*{-1.5mm}
\end{enumerate}
Here the Cases\,(a)-(e) are generic fits, and the rest of cases are
more or less motivated by our fits to the DAMPE CRE data. For illustration,
we summarize the fitting results for ROIs $(0^\circ,30^\circ)$ and $(90^\circ,70^\circ)$
in Table\,\ref{tab:4}, where we define the difference of log-likelihoods
for a given Case-(j) as:
\beqa
-\Delta\!\ln\!{\mathcal L}_{\text{j}}^{}
=\ln[{\mathcal L}(\text{background})] -
 \ln[{\mathcal L}(\text{Case-j})]\,.
\eeqa
Note that a negative value of $\,-\Delta\!\ln\!{\mathcal L}_{\text{j}}^{}$\,
indicates a better fitting quality than the background-only fit
(corresponding to the Case-(a)).
Further discussions are given in Appendix\,\ref{app:A}.
The full results of all the ROIs are summarized in Appendix\,\ref{app:B}.

\begin{table*}[t]
	\centering
	\begin{tabular}{c|r|c|c|c|c|c|c}
		\hline\hline
        \multicolumn{8}{c}{$(\ell,\,b)=(0^\circ\!,\,30^\circ)$}
		\\
		\hline
        Case & $-\Delta\!\ln\!\mathcal{L}$~ &
        \multicolumn{2}{c|}{$e^\pm$ channel} &
        \multicolumn{2}{c|}{$\mu^\pm$ channel} &
        \multicolumn{2}{c}{$\tau^\pm$ channel}
		\\
		\cline{3-8}
		& & best fit & UL & best fit & UL & best fit & UL
		\\
		\hline
		(a) & $0.00$~ & / & / & / & / & / & /
		\\
		\hline
		(b) & $-0.05$~ & 0.16$\pm$0.54 & 1.11 & / & / & / & /
		\\
		\hline
		(c) & $-0.30$~ & / & / & 0.54$\pm$0.72 & 1.79 & / & /
		\\
		\hline
		(d) & $0.00$~ & / & / & / & / & $10^{-7}$ & 0.076
		\\
		\hline
		(e) & $0.00$~ & $10^{-8}$ & 0.071 & \multicolumn{2}{c|}{same as $e^\pm$} & \multicolumn{2}{c}{same as $e^\pm$}
		\\
		\hline
		(f) & $-0.15$~ & 0.154 & / & 0.34$\pm$0.72 & 1.58 & / & /
		\\
		\hline
		(g) & $-0.05$~ & 0.154 & / & / & / & $10^{-7}$ & 0.066
		\\
		\hline
		(h) & $-0.17$~ & 0.154 & / & 0.34$\pm$0.72 & 1.60 & $10^{-7}$ & 0.061
		\\
		\hline
		(i) & $0.79$~ & 0.154 & / & 1.39 & 1.39 & $10^{-8}$ & 0.037
		\\
		\hline\hline
        \multicolumn{8}{c}{$(\ell,\,b)=(90^\circ\!,\,70^\circ)$}
		\\
		\hline
		Case & $-\Delta\!\ln\!{\mathcal L}$ &
		\multicolumn{2}{c|}{$e^\pm$ channel} &
		\multicolumn{2}{c|}{$\mu^\pm$ channel} &
		\multicolumn{2}{c}{$\tau^\pm$ channel}
		\\
		\cline{3-8}
		& & best fit & UL & best fit & UL & best fit & UL
		\\
		\hline
		(a) & $0.00$~ & / & / & / & / & / & /
		\\
		\hline
		(b) & $0.00$~ & $10^{-11}$ & 0.072 & / & / & / & /
		\\
		\hline
		(c) & $0.00$~ & / & / & $10^{-12}$ & 0.087 & / & /
		\\
		\hline
		(d) & $0.00$~ & / & / & / & / & $10^{-11}$ & 0.015
		\\
		\hline
		(e) & $0.00$~ & $10^{-9}$ & 0.011 & \multicolumn{2}{c|}{same as $e^\pm$} & \multicolumn{2}{c}{same as $e^\pm$}
		\\
		\hline
		(f) & $2.93$~ & 0.154 & / & $10^{-10}$ & 0.083 & / & /
		\\
		\hline
		(g) & $2.93$~ & 0.154 & / & / & / & $10^{-11}$ & 0.014
		\\
		\hline
		(h) & $2.93$~ & 0.154 & / & $10^{-10}$ & 0.083 & $10^{-12}$ & 0.014
		\\
		\hline
		(i) & $28.66$~ & 0.154 & / & 1.39 & 1.39 & $10^{-9}$ & 0.011
		\\
		\hline\hline
	\end{tabular}
\vspace*{3mm}
\caption{\small
Likelihood analysis for selected ROIs with
$(\ell,\,b)=(0^\circ,30^\circ)$ and $(90^\circ,70^\circ)$.
Columns from left to right correspond to the model flags,
$-\Delta\!\ln{\mathcal L}$\,,\, the best-fit
and the 95\%\,C.L.\,upper limits (ULs) of the cross
sections (in unit of $10^{-25}$cm$^{3}$\,s$^{-1}$) for each channel, respectively.
}
\label{tab:4}
\vspace*{2mm}
\end{table*}

\vspace*{1mm}

We find that in general there lacks any DM signal from the selected
ROIs. The corresponding upper limits (ULs) on the
thermally averaged annihilation cross sections
are derived to be about
\,$(0.1\!-\!1)\!\times\!10^{-25}$\,cm$^3$\,s$^{-1}$\,
for the $e^{\pm}$ and $\mu^{\pm}$ channels,
and $1\!-\!2$\, orders of magnitude stronger for $\tau^{\pm}$ channel.
These limits depend on the directions
of the sky regions which have different background intensities.
The weakest constraints come from the direction towards the Galactic
center region, where the backgrounds are the highest.
For instance, when the DM subhalo is located at high galactic latitude
such as the ROI $(90^\circ,\,70^\circ)$, the existence of such a
DM subhalo may be marginally constrained by the data.
On the other hand, when the DM subhalo sits
at low galactic latitudes such as the ROI $(0^\circ,\,30^\circ)$,
we see that the annihilation channels $\chi\chi\!\to e^+e^-\!,\mu^+\mu^-$
are still allowed, but as expected, the channel $\chi\chi\!\to\tau^+\tau^-$
is strongly constrained.

\vspace*{1mm}

The Cases\,(f)-(h) which can potentially explain the peak excesses
of the DAMPE measurement are generally consistent with the Fermi-LAT data.
Only in a few ROIs the inclusion of such a DM component leads to
slight tension with the data
($-\Delta\!\ln{\mathcal L}\lesssim 3.5$\,).
The Case-(i) is proposed to explain both the peak excess around
1.4\,TeV and the non-peak feature below 1\,TeV,
which can be further constrained by the Fermi-LAT data.
But, when the subhalo is located in low galactic
latitude regions (such as $b<30^{\circ}$), this case is also consistent
with the Fermi-LAT data. Whenever $\tau^\pm$ channel is open,
it is shown that the $\tau^\pm$ fraction in the
DM annihilation final states is severely constrained.
This also provides important guideline for the DM model building.

\vspace*{1mm}
\subsection{\,Constraints on ${\tau^\pm}$ Fraction}	
\label{sec:tau-ratio}
\label{sec:4.4}
\vspace*{1mm}

In this subsection we consider the flavor ratios
$\,y_\mu^{}\!=\left<\sigma v\right>_\mu^{}\!/\!\left<\sigma v\right>_e^{}\,$ and
$\,y_\tau^{}\!=0.178\left<\sigma v\right>_\tau^{}\!/\!\left<\sigma v\right>_e^{}$.
We scan the parameter space of the full plane of
$\,y_\mu^{}\!- y_\tau^{}$\,
to further pin down the viable region for the flavor ratios in certain ROIs.

\begin{figure*}[t]
\centering
\includegraphics[height=5.8cm]{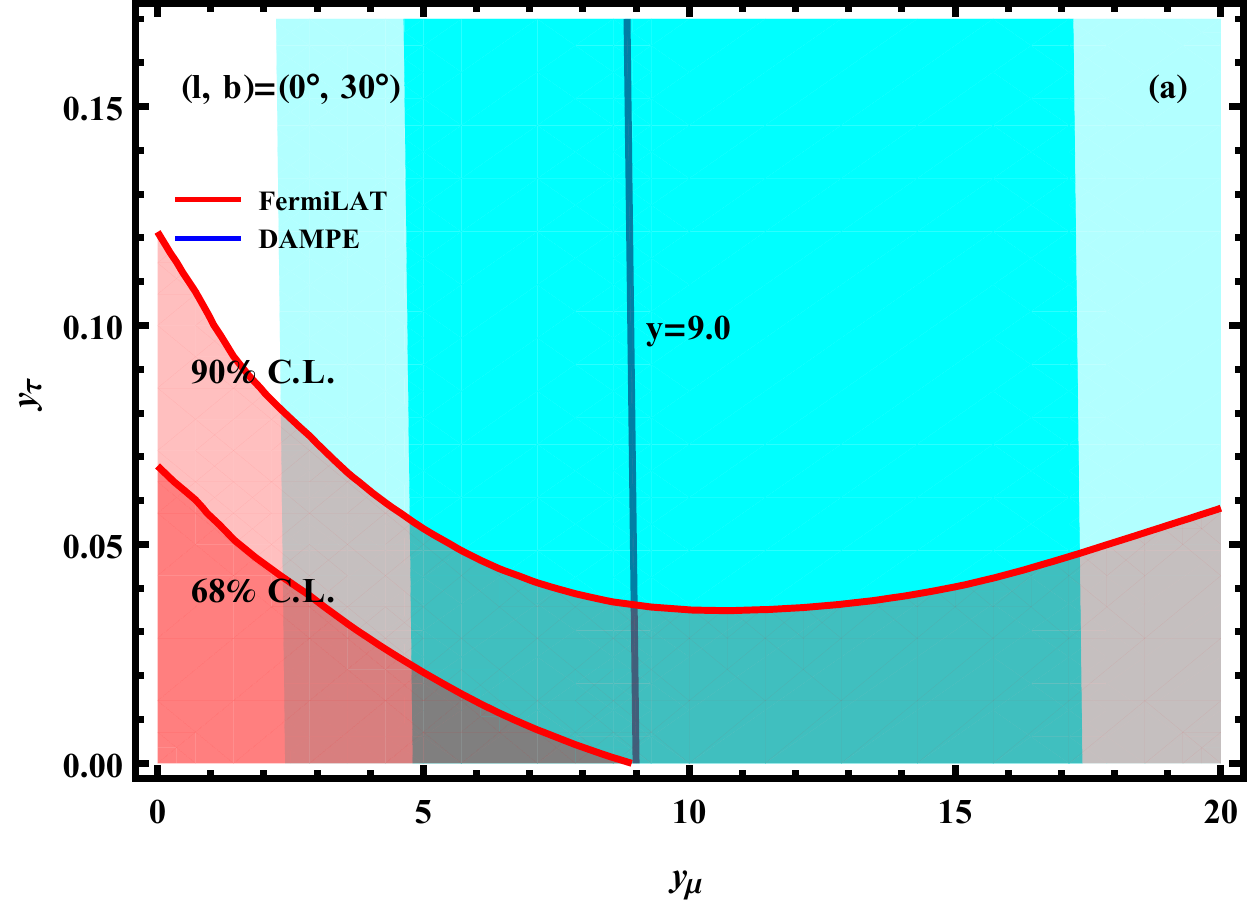}
\hspace*{3mm}
\includegraphics[height=5.8cm]{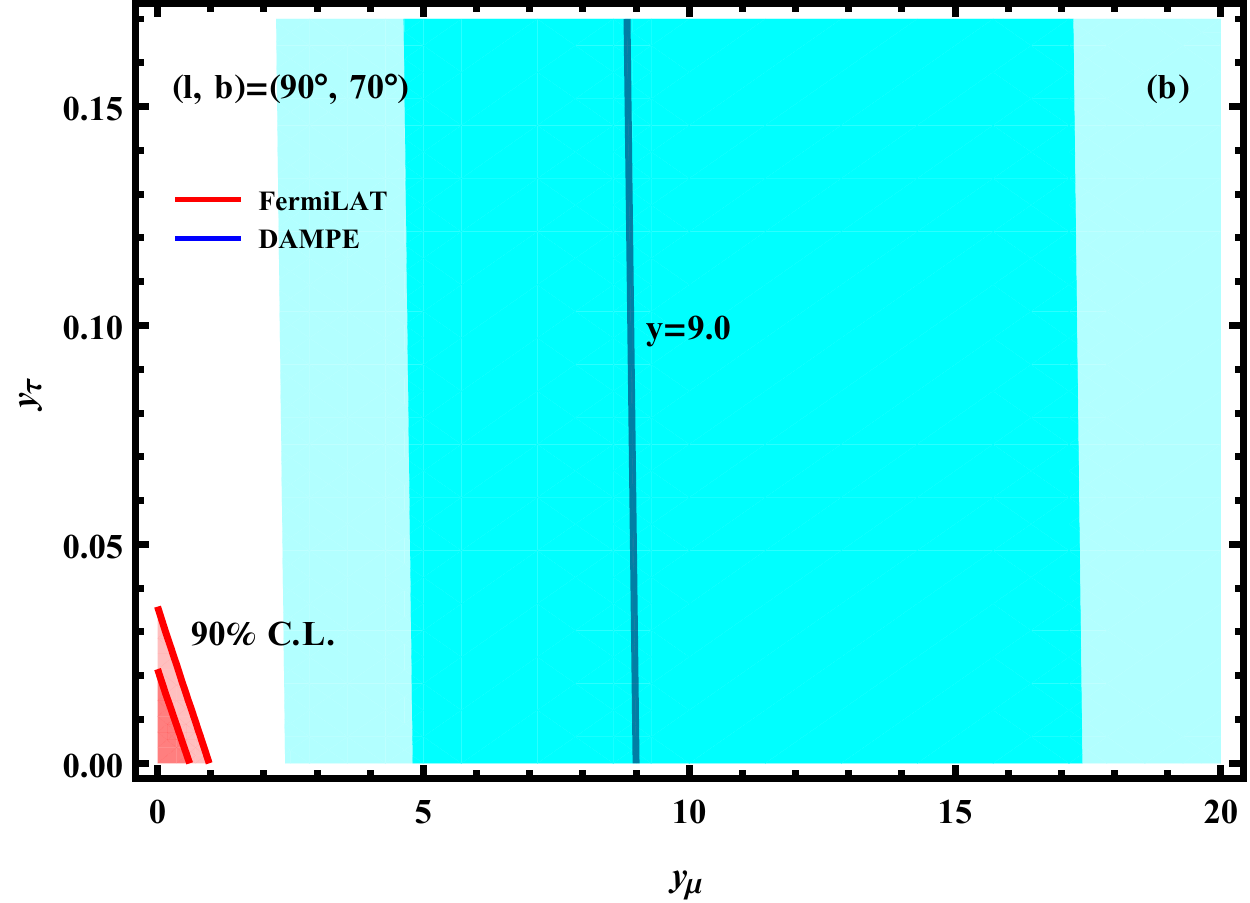}
\vspace*{-2mm}
\caption{\small
Allowed regions in the plane of flavor ratios $y_\mu^{}\!-y_\tau^{}$.\,
In each plot, the red region (pink region) are allowed by the
$\gamma$-ray constraints of Fermi-LAT at 68\% (90\%) confidence limit,
while the blue region (light blue region) are allowed by the DAMPE constraints
at 68\% (90\%) confidence limit.
Plot-(a) is for the ROI centered in $(\ell,\,b)=(0^\circ,\,30^\circ)$
and plot-(b) is for the ROI centered in $(\ell,\,b)=(90^\circ,70^\circ)$.
}
\label{fig:gamma-contours}
\label{fig:9}
\vspace*{-4mm}
\end{figure*}

\vspace*{1mm}

In Fig.\,\ref{fig:9}(a), we plot the result in the ROI with
$(\ell,\,b)=(0^\circ,\,30^\circ)$.\,
We scan the values of $(y_\mu^{},\,y_\tau^{})$ in the regions
$\,y_\mu^{}\!\in\! [0,\,20]$ and $\,y_\tau^{}\!\in\![0,\,0.20]$.
As for the cross sections, we apply the fit relation for
\,$y_\text{total}^{}$ versus $\,\langle\sigma v\rangle_e^{}$\,
as in Fig.\,\ref{fig:3}(d),
rather than using the fixed value
$\langle\sigma v\rangle_e^{}\!=1.54\!\times\!10^{-26}\text{cm}^3\!/\text{s}$\,.\,
For the Fermi-LAT data, the $\chi^2$ function is defined as
$\chi^2(y_\mu^{},y_\tau^{})= -2\ln\!\mathcal{L}$\,,
and $\Delta\chi^2$ is the variation from its best fit point.
In Fig.\,\ref{fig:9}(a), we present the allowed regions by the $\gamma$-ray data
of Fermi-LAT, in the $(y_\mu^{},\,y_\tau^{})$ plane 
at the 68\% (red region) and 90\% (pink region) confidence limits.
In the same $(y_\mu^{},\,y_\tau^{})$ plane, we present the DAMPE constraints from
Fig.\,\ref{fig:3}(b) as the shaded blue region (light blue region) at
the 68\% (90\%) confidence limit.
We see that for $y_\tau^{}=0.035$,\, the case of
$\,N_e^{}\!:\!N_\mu^{}\!:\!N_\tau^{}\simeq 1\!:\!9.0\!:\!0.2$\,
is fully allowed at 90\%\,C.L. in this ROI.
For the lepton portal DM models discussed in Section\,\ref{sec:2.5}, this
corresponds to the coupling relation,
$\,\lambda_e^{}\!:\!\lambda_\mu^{}\!:\!\lambda_\tau^{}
 \!\simeq 1\!:\!1.7\!:\!0.7$\,.

\vspace*{1mm}

Note that as explained in Fig.\,\ref{fig:3}(d) and Section\,\ref{sec:2.4},
for a large ratio $y$\,,\,
the combined contribution from the $\mu^\pm$\,($\tau^\pm$) channels does not
vary much as $\,y\,$ increases,
but the fraction of $e^\pm$ channel will drop.
Thus, the total $\gamma$-ray flux from $e^\pm$ channel will decrease,
which allows more $\tau^\pm$ contribution.
This is why for the 90\%\,C.L.\ contour of Fermi-LAT
in Fig.\,\ref{fig:9}(a), the $\tau^\pm$ fraction rises up in the region of
$\,y_\mu^{}\!\gtrsim\! 10$\,.\,
We see that in the ROI with $(\ell,\,b)=(0^\circ,\,30^\circ)$,\,
the 90\%\,C.L.\ pink region allowed by the Fermi-LAT $\gamma$-ray constraint
has significant overlap with that of the DAMPE CRE constraint.

\vspace*{1mm}

For comparison, we further present the contour plot in Fig.\,\ref{fig:9}(b)
for the ROI with $(\ell,\,b)=(90^\circ,70^\circ)$.
It shows that in this case the Fermi-LAT $\gamma$-ray constraint is so strong
that the DM subhalo scenario in this ROI is already excluded at the 90\%\,C.L.

\vspace*{1mm}
\section{Conclusions}
\label{sec:5}
\vspace*{1mm}

Detecting signals of TeV energy CREs
has been an exciting frontier for many experimental activities in
recent years\,\cite{HESS}-\cite{DAMPE2017}.
These provide important means for probing the possible annihilations 
or decays of dark matter (DM) particles and the nearby galactic sources.

\vspace*{1mm}

In Section\,\ref{sec:2}, we performed a systematically improved
analysis of the DAMPE CRE data\,\cite{DAMPE2017},
with focus on the new hidden excess of non-peak structure over the
energy range $(0.6-\!1.1)$\,TeV (red bins in Fig.\,\ref{fig:1}),
in connection to the peak excess around $(1.3\!-\!1.5)$\,TeV
(blue bin in Fig.\,\ref{fig:1}).
We explained this new non-peak excess and the peak excess from
the 1.5\,TeV $\mu^\pm$\,($\tau^\pm$) events and the 1.5\,TeV $e^\pm$ events
which were produced together from DM annihilations in a nearby subhalo,
with the subsequent 3-body decays of $\mu^\pm$ ($\tau^\pm$) into $e^\pm$ plus
neutrinos (Fig.\,\ref{fig:2}).
We made an improved analysis of the CRE physical backgrounds which
consist of CRE fluxes from SNR and ISM.
Then, we systematically analyzed the CRE spectra from the DM annihilations
including the final state photon radiations (which soften the CRE spectra),
as shown in Fig.\,\ref{fig:2}.
Our improved fit analysis demonstrated that the {\it flavor structure} of the original
lepton final-state of DM annihilations
in a nearby subhalo or clump should have a flavor composition ratio
$N_e \!:\! (N_\mu \!+\!0.178N_\tau) = 1\!:\!y$\,
with $\,y\!=\! 9.0^{+8.4}_{-4.2}$\,
($1\sigma$ bound) and $\,2.4\!<\!y\!<\!34.7$\, ($90\%$\,C.L.)
for the current DAMPE data, as shown in Fig.\,\ref{fig:3}.
Using this new fit, we analyzed the lepton-portal DM models and
deduced a bound on the lepton-DM-mediator couplings
$\,\lambda_e^{} \!: (\lambda_\mu^4 \!+\frac{1}{6}\lambda_\tau^4)^{\frac{1}{4}}
=1\!:y^{\frac{1}{4}}\,$
with a narrow range $\,y^{\frac{1}{4}}\simeq 1.2\!-2.4\,$.
Such constraints provide important guideline for DM model buildings.
We further analyzed the improved sensitivities from the projected 4-year and
6-year runnings of DAMPE detector, as presented in Fig.\,\ref{fig:4}.
We found that by assuming the current central values of the CRE spectrum,
the DAMPE 6-year running can constrain the lepton flavor ratio to
$\,y= 8.0^{+3.9}_{-3.0}$\, at $1\sigma$ level and
$\,3.4<y<15.7$\, at $90\%$\,C.L.,
while the bounds from the 4-year running give
$\,y= 8.4^{+4.6}_{-3.3}$\, at $1\sigma$ level
and $\,3.3<y<18.2$\, at $90\%$\,C.L.
The above systematic analyses fully go beyond our previous
short letter\,\cite{PLB}.

\vspace*{1mm}

In Section\,\ref{sec:3},
extending our analysis with DAMPE data\,\cite{DAMPE2017}
in Section\,\ref{sec:2},
we further combined with the CRE measurement of Fermi-LAT\,\cite{Fermi},
and then compared with the recent data from
AMS-02\,\cite{AMS} and CALET\,\cite{CALET}.
We found that
the new non-peak excess of DAMPE is consistent with the
Fermi-LAT data\,\cite{Fermi}.
The absence of the peak excess ($\sim\!\!1.4$\,TeV) at Fermi-LAT  
may be due to its much lower energy resolution above 1\,TeV scale.
Our combined fit of the DAMPE and Fermi-LAT spectra is presented in
Fig.\,\ref{fig:5} and constrains the lepton flavor ratio as
$\,y= 6.6^{+4.9}_{-2.8}$\, at $1\sigma$ level
and $\,2.2\!<\! y \!<\! 18.1$\, at $90\%$\,C.L.
On the other hand, the data of AMS-02\,\cite{AMS} and CALET\,\cite{CALET}
show more differences from DAMPE, especially for the high energy range
$E\!>\!500$\,GeV.
Nevertheless, we observe that the CALET CRE spectrum shows two broad excesses
around the energy windows $(0.75\!-\!1.2)$\,TeV and $(1.5\!-\!2.0)$\,TeV,
although their statistical uncertainties are still rather sizable.
The first excess is compatible with the non-peak excess in the DAMPE data and
the second excess lies at energies somewhat higher than the DAMPE peak excess.
By inputting the DM mass $M_\chi^{}\!\!=\!1.78$\,TeV and using the DM
cross sections from our best fit of DAMPE data,
we can still fit the CALET data with good quality (Fig.\,\ref{fig:6}).
As compared to the pure background fit, the total $\chi^2$ only has a minor
rise $\,\Delta\chi^2=1.59$\,.

\vspace*{1mm}

In Section\,\ref{sec:4},
we further analyzed the $\gamma$-ray measurements of Fermi-LAT
and derived new constraints on the fluxes of different
final state leptons from the DM annihilations.
Our analysis considered for the first time the spatial extension of
the $\gamma$-ray emission from the subhalo (Fig.\,\ref{fig:8}).
We showed in Tables\,\ref{tab:4}-\ref{tab:8}
that the existence of such a nearby DM subhalo
at high galactic latitude is within the sensitivity reach of Fermi-LAT,
although no significant $\gamma$-ray signals show up yet.
For the low galactic latitude regions such as
the ROI centering at $(\ell,b)=(0^\circ,30^\circ)$,\,
we found in Tables\,\ref{tab:4} and \ref{tab:8}
that the DM annihilation channels
$\chi\chi\!\to\!e^+e^-\!,\mu^+\mu^-$ are still viable, while
the channel $\chi\chi\!\to\!\tau^+\tau^-$ is subject to a strong constraint.
We presented the Fermi-LAT constraints together with the DAMPE CRE fit
on the flavor ratios
$(y_\mu^{},\,y_\tau^{})$ in Figs.\,\ref{fig:9}(a) and \ref{fig:9}(b)
for two sample ROIs with low and high galactic latitudes.
In the ROI centered at $(\ell,b)\!=\!(0^\circ,30^\circ)$,\,
Fig.\,\ref{fig:9}(a) shows that the range of the flavor ratio
$\,y_\tau^{}\!\lesssim\!0.035$\,
is always allowed by the Fermi-LAT $\gamma$-ray data.
For instance, the flavor ratios
$(y_\mu^{},\,y_\tau^{})\!=\!(9.0,\,0.035)$\,
present a viable realization, which correspond to
$N_e^{}\!:\!N_\mu^{}\!:\!N_\tau^{}\!\simeq\!1\!:\!9.0\!:\!0.2$\,
and thus the ratio of DM-lepton-mediator couplings
$\,\lambda_e^{}\!:\!\lambda_\mu^{}\!:\!\lambda_\tau^{}
 \simeq 1\!:\!1.7\!:\!0.7$\,.
Such constraints give important guideline for the lepton-portal type
of DM model buildings.

\vspace*{1mm}

Our current study demonstrates that the non-peak new excess around $(0.6\!-\!1.1)$\,TeV
and the tentative 1.4\,TeV peak structure in the DAMPE CRE spectrum provide
encouraging clues to the possible new physics related to DM annihilations (decays)
and the flavor composition of the final state leptons. 
These deserve more investigations.
The future high energy CRE measurements are truly important for further 
pinning down these intriguing new excesses of events.

\vspace*{6mm}
\noindent
{\bf\large Acknowledgements}
\\[1mm]
HJH and YCW were supported in part by the National NSF of China
(under grants 11675086 and 11835005), by the CAS Center for Excellence
in Particle Physics (CCEPP), and by the National Key R\,\&\,D Program of
China (No.2017YFA0402204). HJH was also supported in part by
the Shanghai Laboratory for Particle Physics and Cosmology under Grant
No.\,11DZ2260700, and by the Office of Science and Technology,
Shanghai Municipal Government (No.\,16DZ2260200).
QY was supported by the National NSF of China (No.\,11722328, 11851305),
and the 100 Talents Program of Chinese Academy of Sciences.




\vspace*{8mm}
\begin{appendix}

\noindent
{\Large\bf Appendix:}

\section{Likelihood and Upper Limits for the ${\gamma}$-Ray Analysis}
\label{app:A}


The log-likelihood function of the model parameter set $\,\boldsymbol{p}$\,
is given by the Possion probability
\begin{eqnarray}
\ln{\mathcal L}(\boldsymbol{p})
\,=\, \sum_{j} (n_j^{}\!\ln e_j^{}\!-e_j^{}\!-\ln n_j^{}!)\,,
\end{eqnarray}
where $n_j^{}$ is the observed number of photons in a given energy and
spatial bin, and $e_j^{}$ is the expected number of photons in the same
bin given the model parameters. The subscript $j$ runs over all energy
and spatial bins. In our analysis, the paremeter set $\boldsymbol{p}$
includes all the background parameters (i.e., spectral parameters of
point sources in the ROI and the normalizations of the diffuse
backgrounds), the cross section and/or the branching fraction of the DM
annihilation. We adopt the profile likelihood method to deal with
the background parameters \cite{Rolke:2004mj}.

\vspace*{1mm}

In case that no signal is found, we derive the upper limits on
the DM annihilation cross sections.
The 95\% upper limit for one-parameter fit is derived by setting
$\,-\Delta\!\ln\!{\mathcal L}=\ln\!{\mathcal L}(\langle\sigma v\rangle\!\!=\!0)-
 \ln\!{\mathcal L}(\langle\sigma v\rangle)< 1.35$\,.\,
For the Case-(h) which has two free parameters of the DM annihilation,
the 95\%\,C.L.\,upper limit is defined by
$\,-\Delta\!\ln\!{\mathcal L}<2.30$\,.\,

\section{Summary of $\gamma$-ray Fitting Results for All ROIs}
\label{app:B}

In this Appendix, we summarize the fitting results in all ROIs,
in addition to Table\,\ref{tab:4} as presented in Section\,\ref{sec:4.3}.
In each table, the columns from left to right correspond to the model flags,
$-\Delta\ln{\mathcal L}$, best-fit and $95\%$ upper limit (UL) of the
cross sections (in units of $10^{-25}$cm$^{3}$\,s$^{-1}$) for each channel.

\vspace*{1mm}

From Tables\,\ref{tab:A1}-\ref{tab:A3},
we see that at high latitudes with $b\!\geqq\!70^\circ$,
the Fermi-LAT $\gamma$-ray data not only show null results for
the DM annihilation signal at its best fits,
but also exclude our model motivated by the DAMPE data at $95\%$ C.L.

\vspace*{1mm}

On the other hand, in the ROIs at low latitudes with $b\!\leqq\!30^\circ$,
we find that our proposed DM subhalo is allowed by the constraints of the
10-year $\gamma$-ray data from Fermi-LAT experiment,
although no significant signals show up yet.
But in this case, we can derive an upper bound
on the flavor ratio of \,$\tau^\pm$.\,
We present this analysis in Table\,\ref{tab:4} (Section\,\ref{sec:4.3})
and the following Table\,\ref{tab:8}.

\begin{table*}[t]
	\centering
	\begin{tabular}{c|r|c|c|c|c|c|c}
		\hline\hline
		Case & $-\Delta\!\ln\!{\mathcal L}$~ & \multicolumn{2}{c|}{$e^\pm$ channel} & \multicolumn{2}{c|}{$\mu^\pm$ channel} & \multicolumn{2}{c}{$\tau^\pm$ channel}
		\\
		\cline{3-8}
		& & best fit & UL & best fit & UL & best fit & UL
		\\
		\hline\hline
		\multicolumn{8}{c}{$(\ell,\,b)=(0^\circ\!,\,90^\circ)$}
		\\
		\hline
		(a) & $0.00$~ & / & / & / & / & / & /
		\\
		\hline
		(b) & $0.00$~ & $10^{-6}$ & 0.143 & / & / & / & /
		\\
		\hline
		(c) & $0.00$~ & / & / & $10^{-6}$ & 0.160 & / & /
		\\
		\hline
		(d) & $0.00$~ & / & / & / & / & $10^{-6}$ & 0.037
		\\
		\hline
		(e) & $0.00$~ & $10^{-6}$ & 0.025 & \multicolumn{2}{c|}{same as $e^\pm$} & \multicolumn{2}{c}{same as $e^\pm$}
		\\
		\hline
		(f) & $1.46$~ & 0.154 & / & $10^{-6}$ & 0.133 & / & /
		\\
		\hline
		(g) & $1.46$~ & 0.154 & / & / & / & $10^{-7}$ & 0.028
		\\
		\hline
		(h) & $1.46$~ & 0.154 & / & $10^{-6}$ & 0.133 & $10^{-7}$ & 0.028
		\\
		\hline
		(i) & $20.52$~ & 0.154 & / & 1.39 & 1.39 & $10^{-7}$ & 0.015
		\\
		\hline\hline
		\multicolumn{8}{c}{$(\ell,\,b)=(0^\circ\!,\,-90^\circ)$}
		\\
		\hline
		(a) & $0.00$~ & / & / & / & / & / & /
		\\
		\hline
		(b) & $0.00$~ & $10^{-6}$ & 0.093 & / & / & / & /
		\\
		\hline
		(c) & $0.00$~ & / & / & $10^{-6}$ & 0.108 & / & /
		\\
		\hline
		(d) & $0.00$~ & / & / & / & / & $10^{-6}$ & 0.011
		\\
		\hline
		(e) & $0.00$~ & $10^{-6}$ & 0.016 & \multicolumn{2}{c|}{same as $e^\pm$} & \multicolumn{2}{c}{same as $e^\pm$}
		\\
		\hline
		(f) & $2.27$~ & 0.154 & / & $10^{-6}$ & 0.101 & / & /
		\\
		\hline
		(g) & $2.27$~ & 0.154 & / & / & / & $10^{-7}$ & 0.021
		\\
		\hline
		(h) & $2.27$~ & 0.154 & / & $10^{-6}$ & 0.100 & $10^{-7}$ & 0.021
		\\
		\hline
		(i) & $24.31$~ & 0.154 & / & 1.39 & 1.39 & $10^{-7}$ & 0.014
		\\
		\hline\hline
	\end{tabular}
\vspace*{3mm}
\caption{\small
The likelihood analysis for ROIs centered at the north and south Galactic poles.
Columns from left to right correspond to the model flags,
$-\Delta\!\ln\!{\mathcal L}$\,,\, best-fit,
and $95\%$ confidence upper limits (UL) of the cross sections (in unit of
$10^{-25}$cm$^{3}$s$^{-1}$) for each channel, respectively.	
}
\label{tab:A1}
\label{tab:5}
\vspace*{3mm}
\end{table*}

\begin{table*}[t]
	\centering
	\begin{tabular}{c|r|c|c|c|c|c|c}
		\hline\hline
		Case & $-\Delta\!\ln\!{\mathcal L}$~ & \multicolumn{2}{c|}{$e^\pm$ channel} & \multicolumn{2}{c|}{$\mu^\pm$ channel} & \multicolumn{2}{c}{$\tau^\pm$ channel}
		\\
		\cline{3-8}
		& & best fit & UL & best fit & UL & best fit & UL
		\\
		\hline\hline
		\multicolumn{8}{c}{$(\ell,\,b)=(0^\circ,70^\circ)$}
		\\
		\hline
		(a) & $0.00$~ & / & / & / & / & / & /
		\\
		\hline
		(b) & $0.00$~ & $10^{-11}$ & 0.066 & / & / & / & /
		\\
		\hline
		(c) & $0.00$~ & / & / & $10^{-7}$ & 0.080 & / & /
		\\
		\hline
		(d) & $0.00$~ & / & / & / & / & $10^{-9}$ & 0.013
		\\
		\hline
		(e) & $0.00$~ & $10^{-11}$ & 0.010 & \multicolumn{2}{c|}{same as $e^\pm$} & \multicolumn{2}{c}{same as $e^\pm$}
		\\
		\hline
		(f) & $3.18$~ & 0.154 & / & $10^{-8}$ & 0.077 & / & /
		\\
		\hline
		(g) & $3.18$~ & 0.154 & / & / & / & $10^{-9}$ & 0.012
		\\
		\hline
		(h) & $3.18$~ & 0.154 & / & $10^{-8}$ & 0.077 & $10^{-9}$ & 0.012
		\\
		\hline
		(i) & $30.08$~ & 0.154 & / & 1.39 & 1.39 & $10^{-10}$ & 0.011
		\\
		\hline\hline
		\multicolumn{8}{c}{$(\ell,\,b)=(90^\circ,70^\circ)$}
		\\
		\hline
		(a) & $0.00$~ & / & / & / & / & / & /
		\\
		\hline
		(b) & $0.00$~ & $10^{-11}$ & 0.072 & / & / & / & /
		\\
		\hline
		(c) & $0.00$~ & / & / & $10^{-12}$ & 0.087 & / & /
		\\
		\hline
		(d) & $0.00$~ & / & / & / & / & $10^{-11}$ & 0.015
		\\
		\hline
		(e) & $0.00$~ & $10^{-9}$ & 0.011 & \multicolumn{2}{c|}{same as $e^\pm$} & \multicolumn{2}{c}{same as $e^\pm$}
		\\
		\hline
		(f) & $2.93$~ & 0.154 & / & $10^{-10}$ & 0.083 & / & /
		\\
		\hline
		(g) & $2.93$~ & 0.154 & / & / & / & $10^{-11}$ & 0.014
		\\
		\hline
		(h) & $2.93$~ & 0.154 & / & $10^{-10}$ & 0.083 & $10^{-12}$ & 0.014
		\\
		\hline
		(i) & $28.66$~ & 0.154 & / & 1.39 & 1.39 & $10^{-9}$ & 0.011
		\\
		\hline\hline
		\multicolumn{8}{c}{$(\ell,\,b)=(180^\circ,70^\circ)$}
		\\
		\hline
		(a) & $0.00$~ & / & / & / & / & / & /
		\\
		\hline
		(b) & $0.00$~ & $10^{-6}$ & 0.085 & / & / & / & /
		\\
		\hline
		(c) & $0.00$~ & / & / & $10^{-6}$ & 0.102 & / & /
		\\
		\hline
		(d) & $0.00$~ & / & / & / & / & $10^{-6}$ & 0.018
		\\
		\hline
		(e) & $0.00$~ & $10^{-6}$ & 0.018 & \multicolumn{2}{c|}{same as $e^\pm$} & \multicolumn{2}{c}{same as $e^\pm$}
		\\
		\hline
		(f) & $2.48$~ & 0.154 & / & $10^{-6}$ & 0.096 & / & /
		\\
		\hline
		(g) & $2.48$~ & 0.154 & / & / & / & $10^{-7}$ & 0.016
		\\
		\hline
		(h) & $2.48$~ & 0.154 & / & $10^{-6}$ & 0.096 & $10^{-7}$ & 0.016
		\\
		\hline
		(i) & $24.98$~ & 0.154 & / & 1.39 & 1.39 & $10^{-7}$ & 0.013
		\\
		\hline\hline
		\multicolumn{8}{c}{$(\ell,\,b)=(270^\circ,70^\circ)$}
		\\
		\hline
		(a) & $0.00$~ & / & / & / & / & / & /
		\\
		\hline
		(b) & $0.00$~ & $10^{-6}$ & 0.083 & / & / & / & /
		\\
		\hline
		(c) & $0.00$~ & / & / & $10^{-6}$ & 0.094 & / & /
		\\
		\hline
		(d) & $0.00$~ & / & / & / & / & $10^{-6}$ & 0.021
		\\
		\hline
		(e) & $0.00$~ & $10^{-6}$ & 0.015 & \multicolumn{2}{c|}{same as $e^\pm$} & \multicolumn{2}{c}{same as $e^\pm$}
		\\
		\hline
		(f) & $2.55$~ & 0.154 & / & $10^{-6}$ & 0.088 & / & /
		\\
		\hline
		(g) & $2.55$~ & 0.154 & / & / & / & $10^{-6}$ & 0.019
		\\
		\hline
		(h) & $2.55$~ & 0.154 & / & $10^{-6}$ & 0.088 & $10^{-6}$ & 0.019
		\\
		\hline
		(i) & $27.31$~ & 0.154 & / & 1.39 & 1.39 & $10^{-6}$ & 0.014
		\\
		\hline\hline
\end{tabular}
\vspace*{3mm}
\caption{\small The likelihood analysis for ROIs at high latitudes with $b=70^\circ$.
All the notations are the same as in Table\,\ref{tab:5}.}
\label{tab:A2}
\label{tab:6}
\vspace*{-2mm}
\end{table*}

\begin{table*}[t]
	\centering
	\begin{tabular}{c|r|c|c|c|c|c|c}
		\hline\hline
		Case & $-\Delta\!\ln\!{\mathcal L}$~ & \multicolumn{2}{c|}{$e^\pm$ channel} & \multicolumn{2}{c|}{$\mu^\pm$ channel} & \multicolumn{2}{c}{$\tau^\pm$ channel}
		\\
		\cline{3-8}
		& & best fit & UL & best fit & UL & best fit & UL
		\\
		\hline\hline
		\multicolumn{8}{c}{$(\ell,\,b)=(0^\circ,-70^\circ)$}
		\\
		\hline
		(a) & $0.00$~ & / & / & / & / & / & /
		\\
		\hline
		(b) & $0.00$~ & $10^{-11}$ & 0.156 & / & / & / & /
		\\
		\hline
		(c) & $0.00$~ & / & / & $10^{-11}$ & 0.197 & / & /
		\\
		\hline
		(d) & $0.00$~ & / & / & / & / & $10^{-12}$ & 0.028
		\\
		\hline
		(e) & $0.00$~ & $10^{-13}$ & 0.021 & \multicolumn{2}{c|}{same as $e^\pm$} & \multicolumn{2}{c}{same as $e^\pm$}
		\\
		\hline
		(f) & $1.33$~ & 0.154 & / & $10^{-10}$ & 0.177 & / & /
		\\
		\hline
		(g) & $1.33$~ & 0.154 & / & / & / & $10^{-13}$ & 0.025
		\\
		\hline
		(h) & $1.33$~ & 0.154 & / & $10^{-10}$ & 0.177 & $10^{-11}$ & 0.025
		\\
		\hline
		(i) & $15.00$~ & 0.154 & / & 1.39 & 1.39 & $10^{-12}$ & 0.017
		\\
		\hline\hline
		\multicolumn{8}{c}{$(\ell,\,b)=(90^\circ,-70^\circ)$}
		\\
		\hline
		(a) & $0.00$~ & / & / & / & / & / & /
		\\
		\hline
		(b) & $0.00$~ & $10^{-8}$ & 0.106 & / & / & / & /
		\\
		\hline
		(c) & $0.00$~ & / & / & $10^{-9}$ & 0.117 & / & /
		\\
		\hline
		(d) & $0.00$~ & / & / & / & / & $10^{-10}$ & 0.032
		\\
		\hline
		(e) & $0.00$~ & $10^{-9}$ & 0.020 & \multicolumn{2}{c|}{same as $e^\pm$} & \multicolumn{2}{c}{same as $e^\pm$}
		\\
		\hline
		(f) & $1.99$~ & 0.154 & / & $10^{-8}$ & 0.109 & / & /
		\\
		\hline
		(g) & $1.99$~ & 0.154 & / & / & / & $10^{-10}$ & 0.028
		\\
		\hline
		(h) & $1.99$~ & 0.154 & / & $10^{-8}$ & 0.109 & $10^{-11}$ & 0.028
		\\
		\hline
		(i) & $22.69$~ & 0.154 & / & 1.39 & 1.39 & $10^{-9}$ & 0.017
		\\
		\hline\hline
		\multicolumn{8}{c}{$(\ell,\,b)=(180^\circ,-70^\circ)$}
		\\
		\hline
		(a) & $0.00$~ & / & / & / & / & / & /
		\\
		\hline
		(b) & $0.00$~ & $10^{-6}$ & 0.106 & / & / & / & /
		\\
		\hline
		(c) & $0.00$~ & / & / & $10^{-6}$ & 0.122 & / & /
		\\
		\hline
		(d) & $0.00$~ & / & / & / & / & $10^{-12}$ & 0.029
		\\
		\hline
		(e) & $0.00$~ & $10^{-9}$ & 0.019 & \multicolumn{2}{c|}{same as $e^\pm$} & \multicolumn{2}{c}{same as $e^\pm$}
		\\
		\hline
		(f) & $1.99$~ & 0.154 & / & $10^{-7}$ & 0.112 & / & /
		\\
		\hline
		(g) & $1.99$~ & 0.154 & / & / & / & $10^{-10}$ & 0.025
		\\
		\hline
		(h) & $1.99$~ & 0.154 & / & $10^{-7}$ & 0.112 & $10^{-10}$ & 0.025
		\\
		\hline
		(i) & $22.41$~ & 0.154 & / & 1.39 & 1.39 & $10^{-10}$ & 0.016
		\\
		\hline\hline
		\multicolumn{8}{c}{$(\ell,\,b)=(270^\circ,-70^\circ)$}
		\\
		\hline
		(a) & $0.00$~ & / & / & / & / & / & /
		\\
		\hline
		(b) & $0.00$~ & $10^{-7}$ & 0.209 & / & / & / & /
		\\
		\hline
		(c) & $0.00$~ & / & / & $10^{-7}$ & 0.243 & / & /
		\\
		\hline
		(d) & $0.00$~ & / & / & / & / & $10^{-13}$ & 0.054
		\\
		\hline
		(e) & $0.00$~ & $10^{-9}$ & 0.037 & \multicolumn{2}{c|}{same as $e^\pm$} & \multicolumn{2}{c}{same as $e^\pm$}
		\\
		\hline
		(f) & $0.92$~ & 0.154 & / & $10^{-7}$ & 0.194 & / & /
		\\
		\hline
		(g) & $0.92$~ & 0.154 & / & / & / & $10^{-10}$ & 0.040
		\\
		\hline
		(h) & $0.92$~ & 0.154 & / & $10^{-7}$ & 0.194 & $10^{-7}$ & 0.040
		\\
		\hline
		(i) & $41.08$~ & 0.154 & / & 1.39 & 1.39 & $10^{-13}$ & 0.013
		\\
		\hline\hline
	\end{tabular}
\vspace*{3mm}
\caption{\small The likelihood analysis for ROIs at high latitutes with
		$b=-70^\circ$. All the notations are the same as in Table\,\ref{tab:5}.}
\label{tab:A3}
\label{tab:7}
\vspace*{-2mm}
\end{table*}

\begin{table*}[t]
	\centering
	\begin{tabular}{c|r|c|c|c|c|c|c}
		\hline\hline
		Case & $-\Delta\!\ln\!\mathcal{L}$~ & \multicolumn{2}{c|}{$e^\pm$ channel} & \multicolumn{2}{c|}{$\mu^\pm$ channel} & \multicolumn{2}{c}{$\tau^\pm$ channel}
		\\
		\cline{3-8}
		& & best fit & UL & best fit & UL & best fit & UL
		\\
		\hline\hline
		\multicolumn{8}{c}{$(\ell,\,b)=(120^\circ,45^\circ)$}
		\\
		\hline
		(a) & $0.00$~ & / & / & / & / & / & /
		\\
		\hline
		(b) & $-0.39$~ & $0.16\pm 0.20$ & 0.541 & / & / & / & /
		\\
		\hline
		(c) & $-0.31$~ & / & / & $0.18\pm 0.25$ & 0.663 & / & /
		\\
		\hline
		(d) & $-0.42$~ & / & / & / & / & $0.024\pm 0.030$ & 0.083
		\\
		\hline
		(e) & $-0.41$~ & $0.019\pm 0.024$ & 0.065 & \multicolumn{2}{c|}{same as $e^\pm$} & \multicolumn{2}{c}{same as $e^\pm$}
		\\
		\hline
		(f) & $-0.39$~ & 0.154 & / & $10^{-6}$ & 0.464 & / & /
		\\
		\hline
		(g) & $-0.39$~ & 0.154 & / & / & / & $10^{-4}$ & 0.061
		\\
		\hline
		(h) & $-0.39$~ & 0.154 & / & $10^{-6}$ & 0.443 & $10^{-4}$ & 0.061
		\\
		\hline
		(i) & $8.79$~ & 0.154 & / & 1.39 & 1.39 & $10^{-9}$ & 0.018
		\\
		\hline\hline
		\multicolumn{8}{c}{$(\ell,\,b)=(0^\circ,30^\circ)$}
		\\
		\hline
		(a) & $0.00$~ & / & / & / & / & / & /
		\\
		\hline
		(b) & $-0.05$~ & 0.16$\pm$0.54 & 1.11 & / & / & / & /
		\\
		\hline
		(c) & $-0.30$~ & / & / & 0.54$\pm$0.72 & 1.79 & / & /
		\\
		\hline
		(d) & $0.00$~ & / & / & / & / & $10^{-7}$ & 0.076
		\\
		\hline
		(e) & $0.00$~ & $10^{-8}$ & 0.071 & \multicolumn{2}{c|}{same as $e^\pm$} & \multicolumn{2}{c}{same as $e^\pm$}
		\\
		\hline
		(f) & $-0.15$~ & 0.154 & / & 0.34$\pm$0.72 & 1.58 & / & /
		\\
		\hline
		(g) & $-0.05$~ & 0.154 & / & / & / & $10^{-7}$ & 0.066
		\\
		\hline
		(h) & $-0.17$~ & 0.154 & / & 0.34$\pm$0.72 & 1.60 & $10^{-7}$ & 0.061
		\\
		\hline
		(i) & $0.79$~ & 0.154 & / & 1.39 & 1.39 & $10^{-8}$ & 0.037
		\\
		\hline\hline
		\multicolumn{8}{c}{$(\ell,\,b)=(0^\circ,0^\circ)$}
		\\
		\hline
		(a) & $0.00$~ & / & / & / & / & / & /
		\\
		\hline
		(b) & $0.00$~ & $10^{-7}$ & 0.803 & / & / & / & /
		\\
		\hline
		(c) & $0.18$~ & / & / & $10^{-7}$ & 1.15 & / & /
		\\
		\hline
		(d) & $0.00$~ & / & / & / & / & $10^{-9}$ & 0.117
		\\
		\hline
		(e) & $0.16$~ & $10^{-7}$ & 0.095 & \multicolumn{2}{c|}{same as $e^\pm$} & \multicolumn{2}{c}{same as $e^\pm$}
		\\
		\hline
		(f) & $0.26$~ & 0.154 & / & $10^{-7}$ & 1.00 & / & /
		\\
		\hline
		(g) & $0.27$~ & 0.154 & / & / & / & $10^{-11}$ & 0.113
		\\
		\hline
		(h) & $0.26$~ & 0.154 & / & $10^{-7}$ & 1.01 & $10^{-4}$ & 0.113
		\\
		\hline
		(i) & $2.20$~ & 0.154 & / & 1.39 & 1.39 & $10^{-10}$ & 0.089
		\\
		\hline\hline
\end{tabular}
\vspace*{3mm}
\caption{\small
The likelihood analysis for ROIs at medium and low Galactic latitudes.
All the notations are the same as in Table\,\ref{tab:5}.
As mentioned in Section\,\ref{sec:4.3},
here the photon energy range for the region $(0^\circ,\,30^\circ)$
is chosen to be $(10\!-\!500)$\,GeV,
instead of the range $(1\!-\!500)$\,GeV in other regions.
}
\label{tab:A4}
\label{tab:8}
\vspace*{-2mm}
\end{table*}
\clearpage

\end{appendix}

\baselineskip 17pt

\newpage

\end{document}